\documentclass[aps,pre,twocolumn,noshowpacs,superscriptaddress,groupedaddress,floatfix]{revtex4}  
\usepackage{graphicx}  
\usepackage{morefloats}
\usepackage{dcolumn}   
\usepackage{bm}        
\usepackage{amssymb}   
\usepackage{color}
\usepackage{times}  
\usepackage{amsmath}
\newcommand{\BigO}[1]{\ensuremath{\operatorname{O}\bigl(#1\bigr)}}
\begin{document}

\widetext
\title{On the behavior of random RNA secondary structures near the glass transition}
\author{William D. Baez}
\affiliation{Department of Physics, The Ohio State University, Columbus, OH 43210}
\author{Kay J\"{o}rg Wiese}
\affiliation{CNRS-Laboratoire de Physique Th\'{e}orique de l'Ecole Normale Sup\'{e}rieure, 24 rue Lhomond, 75005 Paris, France.}
\author{Ralf Bundschuh}
\affiliation{Department of Physics, The Ohio State University, Columbus, OH 43210, USA.}

\date{\today}

\begin{abstract}
RNA forms elaborate secondary structures through intramolecular base pairing.  These structures perform critical biological functions within each cell. Due to the availability of a polynomial algorithm to calculate the partition function over these structures, they are also a suitable  system for the  statistical physics of disordered systems. In this model, below the denaturation temperature, random RNA secondary structures  exist in one of two phases: a strongly disordered, low-temperature glass phase, and a weakly disordered, high-temperature molten phase. The probability of two bases to pair  decays with their distance  with an exponent 3/2 in the molten phase, and about 4/3  in the glass phase. Inspired by previous results from a renormalized field theory of the glass transition separating the two phases, we numerically study this transition. We introduce distinct order parameters for each phase,  that both vanish at the  critical point. We finally explore the driving mechanism behind this transition.
\end{abstract}

\maketitle

\section{Introduction}
Heteropolymer folding is a fundamental biophysical process all living systems rely on. It is of  medical relevance, since, e.g., misfolded proteins and nucleic acids are strongly implicated in the development of several neurological disorders, such as Alzheimer's, Parkinson's, and Lou Gehrig's disease~\cite{pmid21081274, pmid23756188, Taylor2002, pmid28912172, pmid28441058, ISHIGURO2017108}. While general questions of heteropolymer folding can be addressed both in proteins and in ribonucleic acids (RNA), RNA is the simpler of the two systems as it is composed  of only four monomers (as opposed to 20 for proteins), the nucleotides adenine (A), cytosine (C), guanine (G), and uracil (U). These nucleotides have a tendency to form Watson-Crick (A-U, C-G) base pairs. To form these pairs, the RNA strand  folds back onto itself, which leads to the creation of RNA secondary structures. From a  statistical physics standpoint, heteropolymer folding   presents a challenging task for the  physics of disordered systems. In particular, RNA secondary structures are one of the few disordered systems for which one can calculate the partition function in polynomial time~\cite{pmid1695107}. 

Previous studies show that RNA secondary structures exist within one of two well-identified phases. Above a critical temperature $T_{\rm c}$, the system   is in a phase where sequence disorder does not play a significant role. This simplifying assumption allows one to model any random base sequence or a sequence with random pairing energies, $\epsilon_{i,j}$, as a homopolymer. The defining trait for this phase is that the partition function of long RNA molecules is dominated by an exponentially large number of secondary structures with energies differing only at the order of $\BigO{k_{\rm B}T}$. This phase is denoted the {\em molten} phase.

\begin{figure}
\includegraphics[width=\columnwidth]{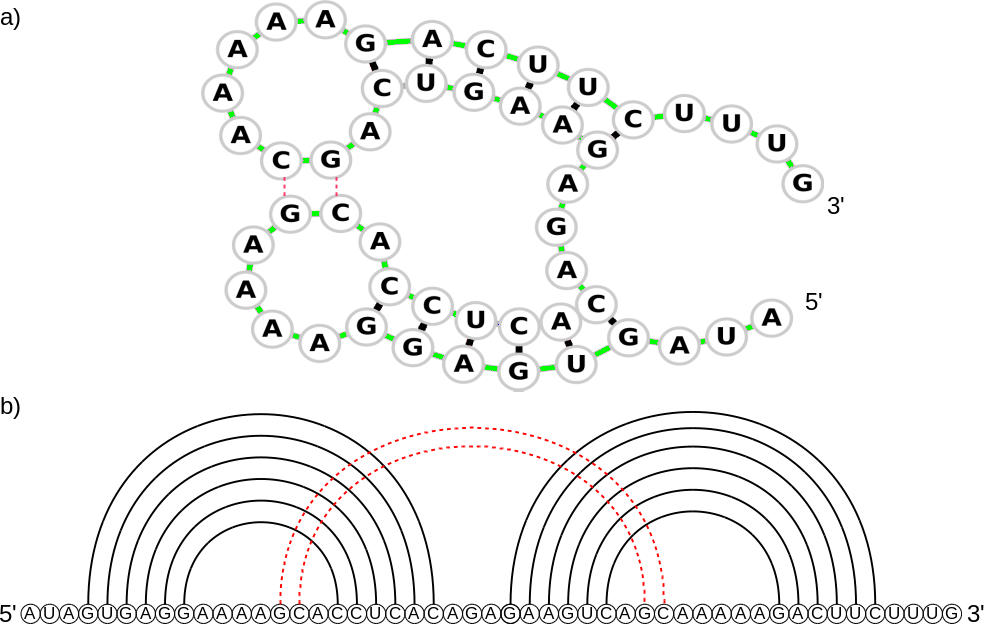}
\caption{\label{fig:secondary_structure} Diagrammatic and abstract representations of an RNA secondary structure. (a) The backbone of the molecule is represented by the solid bright line while the solid black lines stand for the hydrogen-bonded base pairs where each base is depicted as a circle. The shape of the backbone is such that stems of stacked base pairs and the loops connecting or terminating them can be clearly seen. Stems form double-helical structures similar to that of DNA. (b) The same structure as in (a) but with the backbone of the molecule stretched out turning the solid black lines representing each base pair into arches. The base pairs represented by the red dashed lines in (a) and (b) create a pseudoknot visible in (b) as a crossing of the arches. We exclude such pairings in our definition of a secondary structure. Pseudoknots, such as these, do not contribute much to the total energy and are often deemed part of an RNA molecule's tertiary structure.}
\end{figure}
Beginning in 1968, de Gennes~\cite{pgg}, while studying the folded homopolymer,  laid the theoretical foundation for our understanding of the molten phase by showing that the probability of two bases to pair scales as $p(\ell) \sim \ell^{-\alpha}$, where $\ell$ is the distance between any two bases labeled \textit{i} and \textit{j=i+$\ell$} for 1 $\leq$ \textit{i} $<$ \textit{j} $\leq$ N, and  that $\alpha = 3/2$. In later studies~\cite{bund2002pre, bund2002epl} it was demonstrated analytically that sequence disorder, when introduced perturbatively, is irrelevant for small disorder or high temperatures. It was also confirmed numerically that the pairing probability of a four-letter representation of the RNA heteropolymer model matches the scaling behavior of de Gennes' earlier prediction at high enough temperature \cite{bund2002pre, bund2002epl}.

At low temperatures sequence disorder can no longer be ignored. This $\it{glass}$ (or ``frozen'') phase is characterized by the existence of a small number of low-energy secondary structures.  In the glass phase the scaling exponent of the pairing probabilities was found numerically to be $\alpha\approx 1.34$~\cite{muller2002}. Properties of the glass phase have been extensively studied~\cite{muller2002,bund2002pre, bund2002epl, Liu2004, mlkw, fdkw_smallPaper, fdkw_bigPaper, Hui2006, pprt2000, mprt2002, Krzakala2002}. While the existence of the glass phase is well established and the glass phase itself has been characterized extensively, there is yet no clear signature and characterization of the expected phase transition between the molten and the glass phase.

Through the use of a renormalized field theory ~\cite{mlkw}, L\"{a}ssig and Wiese (LW) showed analytically that the freezing transition between the molten and the glass phase is of at least second order. They proved that their model is renormalizable at 1-loop order. They further argued that the value of the exponent $\alpha$ at the transition is the same as in the glass phase. The field theory was later refined by David and Wiese~\cite{fdkw_smallPaper, fdkw_bigPaper}, showing that it is renormalizable to all orders in perturbation theory. At 2-loop order \cite{fdkw_smallPaper, fdkw_bigPaper} it then predicts that $\alpha=3/2$ in the molten phase and $\alpha\approx1.36$ at the transition and in the glass phase.

However, questions concerning the location of this critical point, the existence and behavior of a suitable order parameter, and the transition's driving mechanism remained open. In the present study, through the use of numerical simulations and analytical tools inspired by the field theory, we study two order parameters which both vanish at the transition. One of them is non-zero in the molten phase, the other in the frozen phase. This allows us to precisely locate the critical point. 
Measuring pairing probability distributions,  we explore the mechanism driving this transition.

This paper is organized as follows: In Sec.~\ref{sec:model} we define the model. Sec.~\ref{sec:tcest} details our efforts to locate the critical temperature by establishing two order parameters.  Sec.~\ref{sec:mechanism} probes the transition mechanism. In Sec.~\ref{sec:discussion} we discuss our results.

\section{RNA Secondary-Structure Model} 
\label{sec:model}

\subsection{RNA Secondary Structures}
Each RNA molecule is described by its sequence of $N$ bases $b_1\ldots b_N$, its {\em primary structure}. Given a molecule, an RNA {\em secondary structure}, such as the one shown in Fig.~\ref{fig:secondary_structure}, can then be defined by a list of  ordered pairs $(i,j)$ with $i<j$ representing the base pairs formed by the molecule. Following previous studies in the field of RNA secondary structure reviewed in~\cite{pmid11191843}, we require each base to pair with no more than one other base and we only consider structures that exclude {\em pseudo-knots}. The latter is accomplished by requiring that any two base pairs $(i, j)$ and $(k,l)$ with $i<k$ satisfy either $i < j < k < l $ or $i < k < l < j$ (see Fig.~\ref{fig:secondary_structure}). Configurations violating this rule are termed {\em pseudo-knots}.

While pseudo-knots are biologically relevant~\cite{Chen2005, Staple2005}, the no pseudo-knot constraint is necessary to make both analytical and numerical calculations feasible. The error introduced by the no pseudo-knot constraint is limited due to their relatively infrequent occurrence in real folded RNA~\cite{pmid10550208}. We consider any pseudo-knots and base triples as parts of the {\em tertiary structure} of the molecule, i.e. its 3-dimensional conformation, given the secondary structure. Importantly, we also ignore any other constraints on the secondary strutures due to the tertiary structure, such as, e.g., excluded-volume constraints, in our model.

\subsection{Energy Model}\label{energy-def}
In order to fully describe the statistical physics of RNA secondary structures we have to complement the definition of valid secondary structures by an energy function that assigns an energy to every structure.  While very sophisticated energy models are available~\cite{pmid22115189}, which allow detailed quantitative descriptions of the folding of actual RNA molecules, in this paper we   follow previous studies~\cite{Higgs1996,bund2002pre} that focus  on the universal aspects of RNA secondary structure formation and adopt a simplified RNA energy model. Specifically, we  consider contributions from each base pair, and associate with the pair of  bases labeled
$i$ and $j$ an interaction energy, $\epsilon_{i,j}$. The total energy of a structure $S$ is defined as
\begin{equation}
E[S] = \sum\limits_{(i,j) \in S}  \epsilon_{i,j}.
\end{equation} 
In realistic energy models used in actual secondary-structure prediction of biologically relevant RNA molecules~\cite{pmid22115189}, the interaction energies capture several physically distinct effects:    the enthalpic contributions of hydrogen bonding between the two bases of the pair, stacking between the aromatic rings of two consecutive bases along the strand, and (screened) Coulomb repulsion between the negatively charged phosphate groups of the backbones. In addition, these interaction ``energies'' are  in reality {\em free} energy differences between a paired and an unpaired state. They include the difference in entropy between two unpaired bases freely fluctuating in three-dimensional space and a base pair constrained in its three-dimensional fluctuations due to the requirement to remain paired and due to the greater stiffness of double-stranded   over single-stranded RNA. The enthalpic and entropic contributions nearly balance. At room or body temperature, the enthaltpic contributions slightly outweigh the entropic contributions and RNA forms secondary structures.  At temperatures of around 80$^o$C (depending on the sequence) the entropic contributions outweigh the enthalpic contributions, the interaction free energies become positive, and the RNA   denaturates. Thus, in principle, the interaction (free) energies $\epsilon_{i,j}$ depend on temperature and the identities of the bases $b_i$ and $b_j$ rendering the $\epsilon_{i,j}$ temperature dependent random variables with discrete values and a complicated correlation structure if the RNA sequences are chosen randomly.  Since we are here interested in universal properties of phase transitions, we further simplify our model~\cite{bund2002pre} by choosing these interaction energies as independent Gaussian random variables taken from the distribution
\begin{equation}
\rho(\epsilon) = \frac{1}{\sqrt{2 \pi D}}e^{-\frac{\epsilon^{2}}{2D}}
\end{equation}
with mean energy zero and variance $D$. A mean energy of zero is still low enough to guarantee that most bases engage in base pairs and thus to avoid the denatured phase that we are not interested in.  Also, while we ignore the explicit temperature dependence of the realistic interaction free energies, the universal properties of the phase transition should not depend on which parameter is tuned to cross the transition.

Since $\sqrt{D}$ is the only energy scale in this model, we set $D=1$ in our simulations, but report all results in units scaled by $\sqrt{D}$. 
Finally note that this kind of uncorrelated Gaussian-disorder model is expected to have less severe finite-size effects~\cite{muller2002} than other choices of disorder models.

\subsection{Partition Function}
\label{subsec:partfunc}
Once an energy has been assigned to each secondary structure $S$, the partition function is defined as 
\begin{equation}
Z(N) = \sum\limits_{S \in \Omega(N)}  e^{-\beta E[S]}\ ,
\end{equation}
where $\Omega(N)$ is the set of allowed secondary structures of $N$ bases, and $\beta$ = 1/$k_{\rm B}T$. In the absence of pseudo-knots, the partition function can be studied by considering substrands of the total sequence from base \textit{i} to base \textit{j}. The restricted partition function $Z_{i,j}$ for these substrands obeys the recursive equation~\cite{pmid1695107} 
\begin{equation}\label{eq:recursion}
	Z_{i,j} = Z_{i,j-1} + \sum\limits_{k=i}^{j-1} Z_{i,k-1} \cdot e^{-\beta \epsilon_{k,j}} \cdot Z_{k+1,j-1},
\end{equation}
in which by convention $Z_{i,i-1}=1$.  In this recursion, the right-hand side involves only restricted partition functions for shorter substrands than the one on the left-hand side. Thus, the total partition function, $Z(N) = Z_{1,N}$, can be calculated by progressing from the shortest substrands to longer ones in $\BigO{N^3}$ time. 

The analogous recursion equation~\cite{nussinov1980}
\begin{equation}\label{eq:gs_eq}
E_{i,j} = \min_{i\leq k < j}\left[E_{i,j-1},E_{i,k-1}+\epsilon_{k,j}+E_{k+1,j-1}\right]
\end{equation}
holds for the energy $E_{i,j}$ of the lowest-energy structure on the substrand of the total sequence from base $i$ to base $j$, where $E_{i,i-1}=0$ by convention. This equation can be used in conjunction with a backtracking mechanism~\cite{nussinov1980} to calculate the (zero-temperature) ground-state secondary structure for any disorder configuration $\{\epsilon_{i,j}\}_{i,j}$ in $\BigO{N^3}$ time.

\subsection{Observables}\label{subsec:observables}

{From} the partition function of RNA secondary structures, we can calculate a variety of physical observables that characterize the structural ensemble. Primarily, we study the pairing probability, $p_{i,j}$, for a given base pair $(i,j)$. This probability can be obtained as 
\begin{equation}\label{eq:pij}
p_{i,j} \equiv \frac{e^{-\beta\epsilon_{i,j}}Z_{i+1,j-1}Z_{j+1,i-1}}{Z_{1,N}}\ ,
\end{equation}
where $Z_{i+1,j-1}$ can be calculated from Eq.~(\ref{eq:recursion}) and $Z_{j+1,i-1}$ is the partition function of the sequence $b_{j+1}b_{j+2}...b_{N}b_{1}...b_{i-2}b_{i-1}$. This last quantity can be obtained when the recursion  (\ref{eq:recursion}) is applied to a duplicated sequence $b_{1}...b_{N}b_{1}...b_{N}$ and calculated as $Z_{j+1,N+i-1}$.

We are specifically interested in the dependence of moments of the ensemble averaged base-pairing probability on the distance $\vert i-j\vert=\ell$ between the two bases.  Within the molten phase, and for large $\ell$, this quantity has a power-law dependence
\begin{equation}\label{eq:pbindscaling}
	\langle p(\ell)^{n} \rangle \equiv \langle p_{i,i+\ell}^{n} \rangle\sim \left[\frac{\ell \cdot (N-\ell)}{N}\right]^{-\alpha_n}\ ,
\end{equation}
where the brackets stand for the ensemble average over the random base-pairing energies $\{\epsilon_{i,j}\}_{i,j}$ and $\alpha_{n}$ is a critical exponent. The parameter $n$ allows probing different moments of the distribution of the base pairing probabilities $p(\ell)$. Note that the form (\ref{eq:pbindscaling}) has a practical advantage over the  asymptotic behavior $\ell^{-\alpha_n}$, as it allows us to use larger $\ell$ in extrapolations. 

The logarithms of each pairing probability, 
\begin{equation}
	\Delta F_{i,j}=-k_{\rm B}T\ln(p_{i,j})
\end{equation}
have been interpreted as the ``pinching free energy''~\cite{bund2002pre}, which is the free-energy difference between a pinch of the monomers labeled \textit{i} and \textit{j}, and the unperturbed, or unpinched, state. In Sec.~\ref{sec:tcest}, as in previous studies~\cite{bund2002pre}, we will show how we can use the free energy of the largest possible pinch, 
\begin{equation}
	\Delta F(N) \equiv \Delta F_{1,\frac{N}{2}+1},
\end{equation}
to obtain an estimate of the critical temperature.
Here monomers $1$ and $\frac{N}{2}+1$ are treated as representatives of all splits of a molecule of $N$ bases into two equal pieces of length $\frac{N}{2}-1$. 

\subsection{Review of main results of the L\"assig-Wiese field theory}
\label{subsec:lwreview}

Our numerical analysis of the glass transition is inspired by a field-theoretical calculation by L\"assig and Wiese~\cite{mlkw}, and its refinements in \cite{fdkw_smallPaper, fdkw_bigPaper}.  This theory  starts from the Gaussian-disorder model defined in section \ref{energy-def}. It introduces replicas (copies) of the RNA molecule and shows that the disorder-averaged partition function of the replicated system can be written in terms of two relevant operators. The first is the contact field $\Phi_\alpha(i,j)$ that is $1$ if the bases labeled $i$ and $j$ in replica $\alpha$ are paired and $0$ otherwise.  The second is the overlap field $\Psi_{\alpha,\beta}(i,j)$ between two replicas $\alpha$ and $\beta$ defined as $\Psi_{\alpha,\beta}(i,j)\equiv\Phi_\alpha(i,j)\Phi_\beta(i,j)$, i.e., $\Psi_{\alpha,\beta}(i,j)$ is $1$ if the bases labeled $i$ and $j$ are paired in {\em both} replicas $\alpha$ and $\beta$ and $0$ otherwise.  The scaling dimensions of these operators, i.e., the exponents with which their thermal and disorder-averaged expectation values depend on the distance $|j-i|$ between the two bases, are called $\rho$ and $\theta$, respectively.

\begin{figure}
(a)\hspace*{-4mm}\includegraphics[width=0.5\columnwidth]{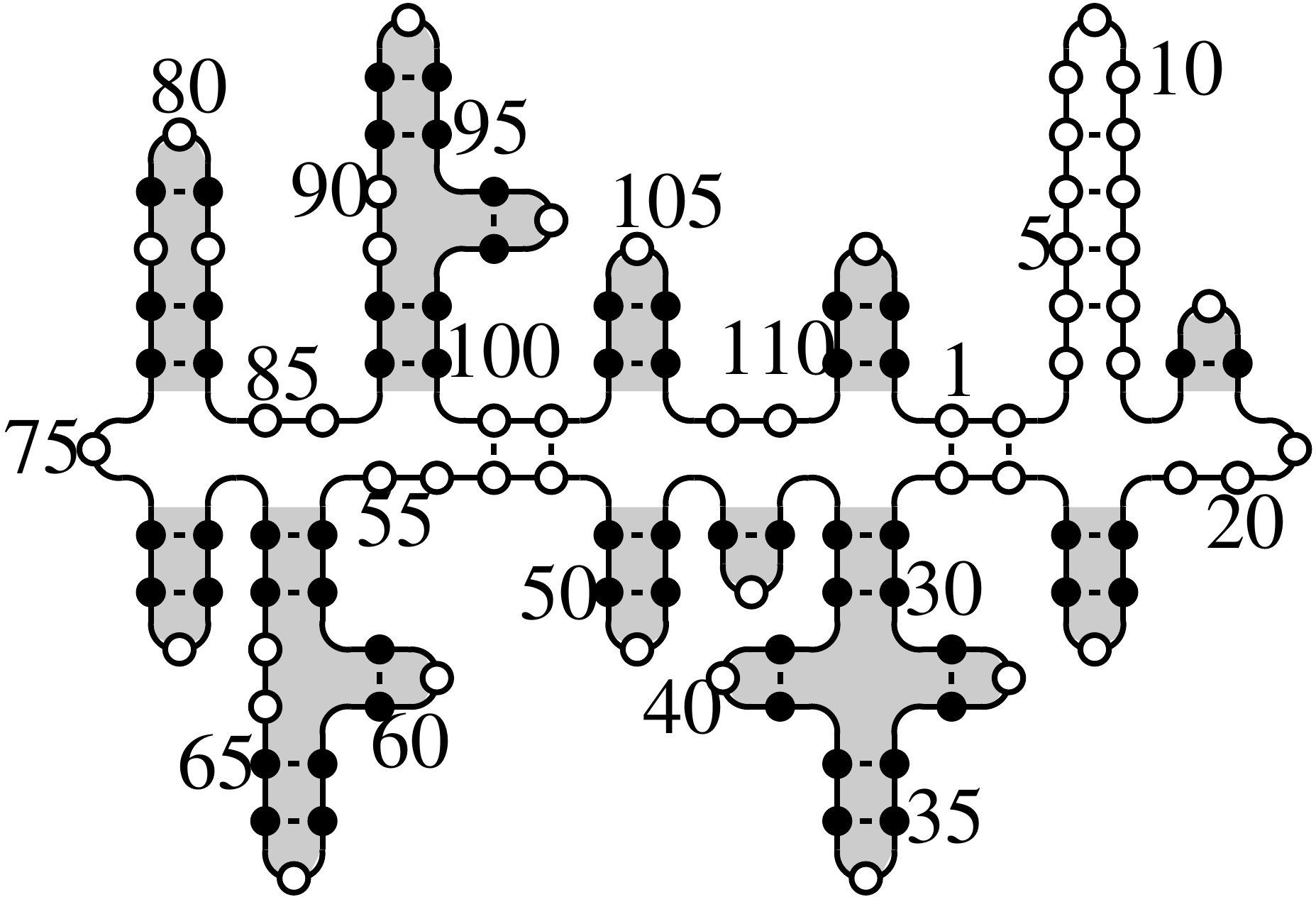}\includegraphics[width=0.5\columnwidth]{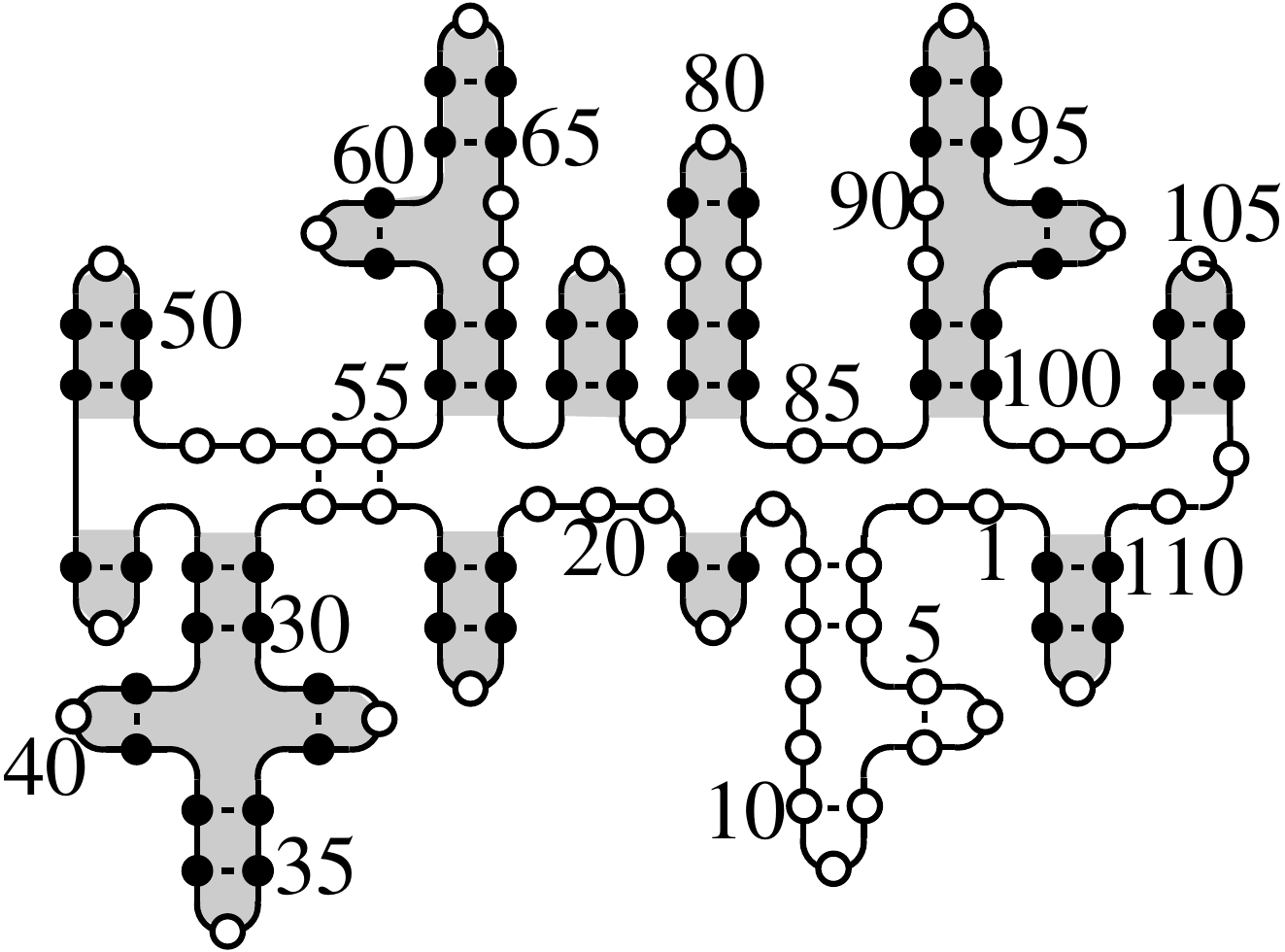}
\\
\rule{\columnwidth}{.1mm}
\vspace{-3mm}
\\
(b)\hspace*{-4mm}\includegraphics[width=0.5\columnwidth]{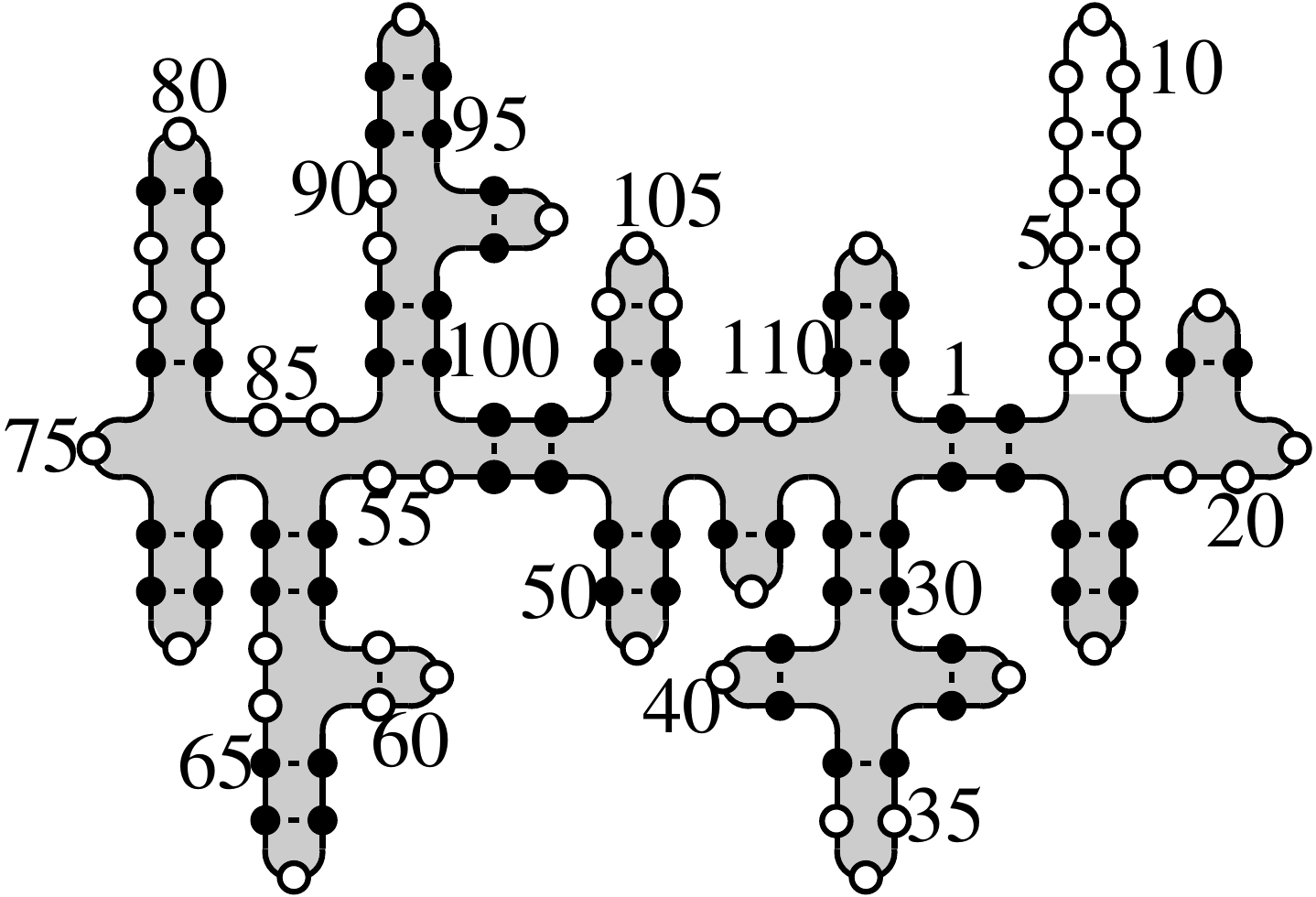}\includegraphics[width=0.5\columnwidth]{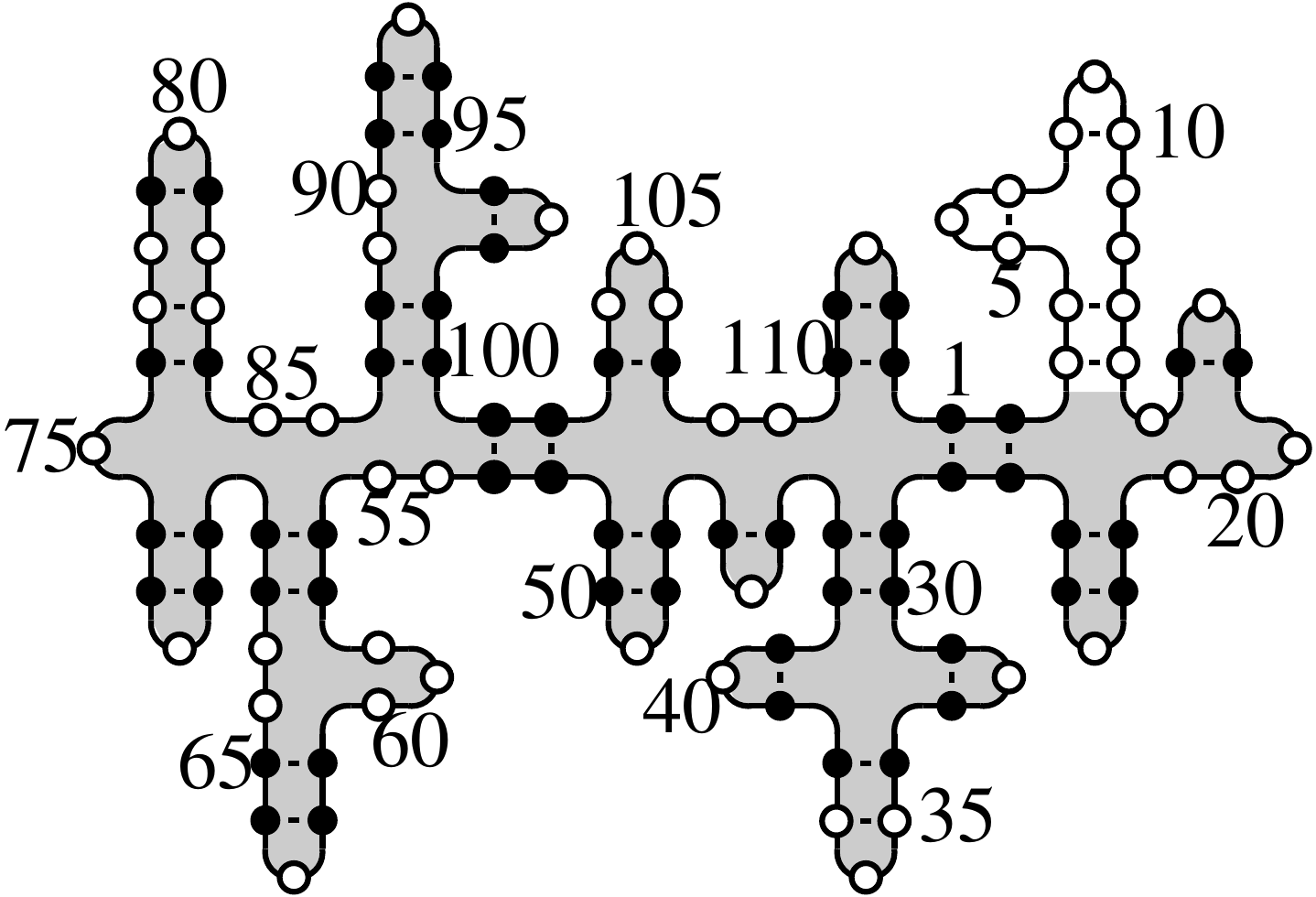}
\caption{Secondary structures of a random RNA molecule at distant times as presented in~\protect\cite{mlkw}. The pairing overlap is defined by the common base pairings between the left and right configuration (corresponding bases are shown in black). (a) Above $T_{\rm c}$, the molecule contains conserved subfolds on scales up to the correlation length $\xi$ (indicated by shading) and is molten on large scales. (b) Below $T_{\rm c}$, the molecule is ``locked'' into its minimum-energy structure on all scales, up to rare fluctuations (unshaded).}
\label{fig:locking_figure}
\end{figure}
L\"assig and Wiese treat the disorder in the model perturbatively, prove renormalizability of the theory at first order (later extended to renormalizability to all orders by David and Wiese~\cite{fdkw_smallPaper, fdkw_bigPaper}), calculate the $\beta$ function, find a critical point that they identify with the glass transition, and calculate the critical exponents $\rho^*$ and $\theta^*$ (the scaling dimensions of $\Phi$ and $\Psi$) at this transition. Importantly, they find expressions for the two critical exponents as a function of the scaling dimension of the disorder and the number of replicas.  In the relevant limit of zero replicas, the expression for the exponent $\theta^*$ (the scaling dimension of $\Psi$ at the transition) becomes smaller than the expression for the exponent $\rho^*$ (the scaling dimension of $\Phi$ at the transition) as the scaling dimension of the disorder approaches its physical value.  Since $\Psi_{\alpha,\beta}(i,j)\le\Phi_\alpha(i,j)$, it is impossible that the probability for two  bases  to be paired in multiple replicas decays slower than the probability for the same two bases  to be paired in a single replica.  Thus, L\"assig and Wiese conclude that rather than crossing each other, the two exponents must become identical, i.e., that $\theta^*=\rho^*$.  They interpret this physically as the different replicas ``locking'' with each other, so that at the transition (and below) there is no difference between looking at the pairing behavior of one or multiple replicas.  This results in the picture of small-scale locked regions in a background of an overall molten structure above the transition temperature, and an overall locked structure with small-scale molten regions below the transition temperature as indicated in Fig.~\ref{fig:locking_figure}.  The length scale $\xi$ of these small-scale regions diverges as $\xi\sim|T-T_{\rm c}|^{-{\nu^*}}$ with $\nu^*=1/(2-\theta^*)$ as the phase transition is approached from either side.  The actual values of the exponents from the 2-loop calculation by David and Wiese~\cite{fdkw_smallPaper, fdkw_bigPaper} are
\begin{equation}\label{eq:moltalphalwvalue}
\theta^*=\rho^*\approx1.36
\end{equation}
and thus
\begin{equation}\label{eq:nuvalue}
\nu^*\approx1.56\ .
\end{equation}
They are very close to the numbers of $\theta^*=\rho^*\approx11/8=1.375$ and $\nu^*\approx8/5=1.6$ provided by L\"assig and Wiese in their original one-loop calculation~\cite{mlkw}.

To connect the results of the L\"assig-Wiese theory to the observables described in section~\ref{subsec:observables}, we have to realize that the pairing probability $p(\ell)$ is the thermal average of the contact field $\Phi_\alpha(i,i+\ell)$.  Since $\rho$ is the scaling dimension of the thermal {\em and} ensemble averaged contact field $\Phi_\alpha(i,i+\ell)$, it describes the scaling of the disorder average of the pairing probability $p(\ell)$, i.e., the quantity described by $n=1$ in Eq.~(\ref{eq:pbindscaling}). Thus,
\begin{equation}\label{eq:rhoconnection}
\alpha_1 = \rho.
\end{equation}
Similarly, the thermal average of the overlap field $\Psi_{\alpha,\beta}(i,i+\ell)$ equals the square of the pairing probability $p(\ell)$, since the bases labeled $i$ and $i+\ell$ have to independently pair in the two replicas.  Since $\theta$ is the scaling dimension of the thermal and ensemble averaged overlap field $\Psi_{\alpha,\beta}(i,i+\ell)$, it describes the scaling of the disorder average of the square of the pairing probability $p(\ell)$, i.e., the quantity described by $n=2$ in Eq.~(\ref{eq:pbindscaling}). Thus,
\begin{equation}\label{eq:thetaconnection}
\alpha_2 = \theta.
\end{equation}

\subsection{Numerical Approach}

To investigate the RNA secondary-structure glass transition, we use RNA molecules of lengths within the range of $500\leq N\leq4000$.  For each length, many independent realizations of the random base-pairing energies $\epsilon_{i,j}$ are chosen, and the partition function of all allowable secondary structures is calculated for each realization of the random variables using Eq.~(\ref{eq:recursion}).  Then, the observables described above are calculated for each realization of the random variables and averaged over those realizations (as well as over the starting base $i$ in case of the base-pairing probability). In the aim to keep the numerical effort manageable we averaged over 20,000 samples for $N=500$, over 10,000 samples for $N\in{750, 1000, 1500}$, over 5,000 samples for $N=2000$, and over 1,000 samples for $N=4000$. We   varied the temperature in a range of  $0.1 \leq k_{\rm B}T/\sqrt{D} \leq 1.0$ to capture the behavior deep within each phase and close to the phase transition, which we will find to be located at $k_{\rm B}T/\sqrt{D}\approx 0.53$.
Note that the smaller number of samples for large systems is partially compensated by a factor of $N$ in the statistics; e.g.\ the number of starting points for base pairs at a given distance grows as $N$, and we average over these starting points. As a result, our statistics are almost independent of size at the chosen numbers of samples.


\section{Scaling of the contact and overlap observables}\label{sec:tcest} 
To understand the model's behavior at the transition, we must know its precise location. Previous studies~\cite{bund2002pre, pprt2000, JankeWieseUnpublished} using different disorder models found that for the system sizes achievable in numerical simulations, thermodynamic signatures of the phase transition are quite weak, making an exact localization of the critical temperature $T_{\rm c}$ challenging. In this section we will identify two order parameters inspired by the L\"assig-Wiese field theory to obtain the precise location of the critical temperature $T_{\rm c}$.

\subsection{Pairing-probability exponents alone do not specify transition temperature}

\begin{figure}[t]
\includegraphics[width=\columnwidth]{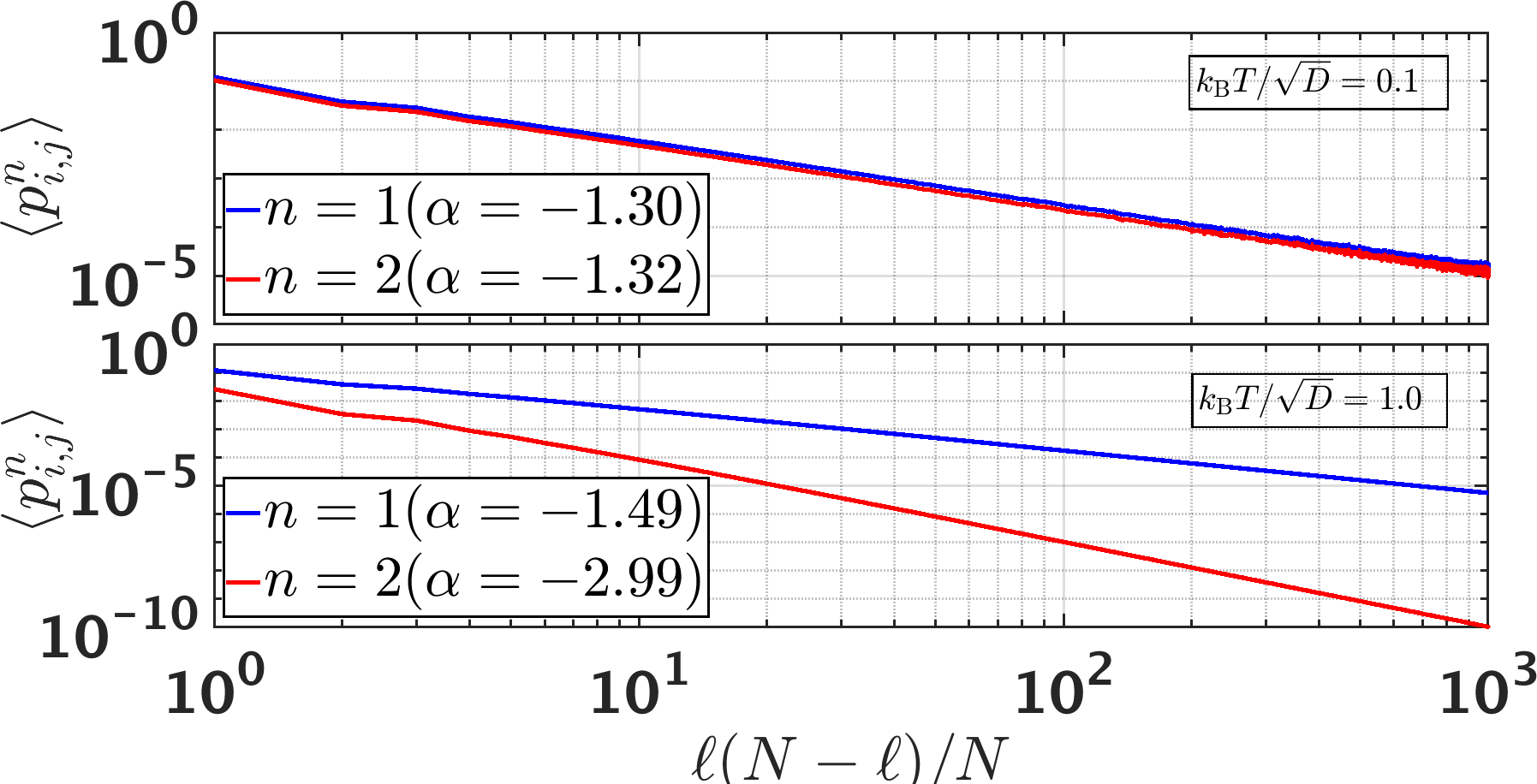}
\caption{\label{fig:powerlawbehavior} Average pairing probability versus substrand length. The upper figure shows how the pairing probability scales as Eq.~(\ref{eq:pbindscaling}) in the glass phase at the lowest temperature considered, $k_{\rm B}T/\sqrt{D}=0.1$. As mentioned in the text, within this phase the critical exponent, $\alpha_n^{(g)}$, is insensitive to the value of $n$. The lower figure shows how the pairing probability scales as Eq.~(\ref{eq:pbindscaling}) in the molten phase at the highest temperature considered, $k_{\rm B}T/\sqrt{D}=1.0$, but demonstrates how the value of $\alpha_n^{(m)}$ directly depends on the value of $n$.\label{fig:alphaathighandlowt}}
\end{figure}
In principle, the critical exponents $\alpha_n$ alone should allow us to locate the phase-transition temperature $T_{\rm c}$.  In the molten phase $p(\ell)$ is independent of disorder and thus $\langle p(\ell)^n \rangle \approx \langle p(\ell) \rangle^n$, i.e.,
\begin{equation}\label{eq:moltenalphavalues}
\alpha_n^{(m)}=n\alpha_1^{(m)}=n\frac{3}{2},
\end{equation}
where we identify exponents in the molten phase with a superscript $(m)$. In the glass phase and at the transition, the L\"assig-Wiese theory predicts (see Eqs.~(\ref{eq:moltalphalwvalue}), (\ref{eq:rhoconnection}), and~(\ref{eq:thetaconnection}))
\begin{equation}\label{eq:glassalphavalues}
\alpha_n^{*}=\alpha_1^{*}\approx1.36\qquad\mbox{and}\qquad
\alpha_n^{(g)}=\alpha_1^{(g)}\approx1.36
\end{equation}
independent of $n$, where we denote exponents at the phase transition with a superscript asterisk and exponents in the glass phase with a superscript $(g)$. 

To verify this, we calculated the disorder-averaged base-pairing probabilities $\langle p(\ell)^{n}\rangle$ numerically for different temperatures at $N=4000$ and fitted them to Eq.~(\ref{eq:pbindscaling}) to obtain $\alpha_n$. Indeed, Fig.~\ref{fig:alphaathighandlowt} shows that the expected power laws hold, for values of $n=1,2$,  both at  low temperature ($T=0.1$)  and at  high temperature ($T=1$).

\begin{figure}[t]
\includegraphics[width=\columnwidth]{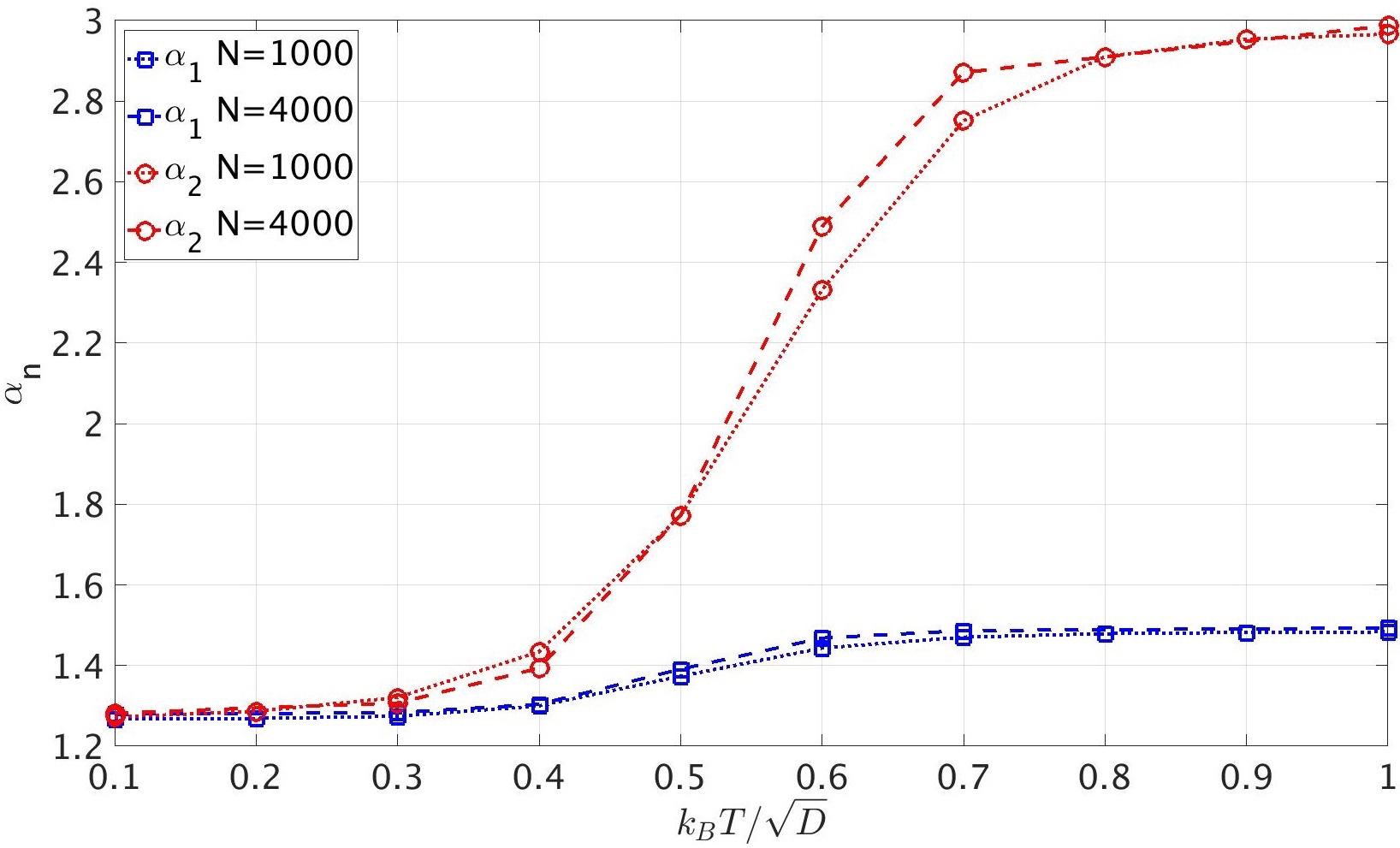}
\caption{The exponents $\alpha_1$ and $\alpha_{2}$ as a function of temperature. As expected, $\alpha_n$ becomes independent of $n$ at a value of $\alpha_n^{(g)}\approx1.36$ at very low temperatures and approaches $\alpha_n^{(m)}=n\frac{3}{2}$ at high temperatures.  However, the changes in apparent $\alpha_n$ are so gradual and the steepness increases only so slightly as the system size is quadrupled that this quantity can neither be used to determine the critical temperature nor the exponents at the transition.}
\label{fig:alphas}
\end{figure}

On the other hand, Fig.~\ref{fig:alphas} shows the behavior of $\alpha_1$ and $\alpha_2$ across the transition, which we show later is at $T_{\rm c} = 0.53\sqrt{D}/k_{\rm B}$. One can see that the apparent values of the exponents change  gradually from their value in the glass phase to their value in the molten phase  as the system undergoes the phase transition. The steepness of this change only increases slightly as the system size is increased from $N=1000$ to $N=4000$. Thus, these exponents cannot be used to precisely locate the critical temperature or to obtain the values of the exponents at the transition.

This gradual change in exponents is consistent with the weakness of the transition observed in earlier work by Pagnani~\textit{et al.}~\cite{pprt2000}. There, it was found that the specific heat has a very broad ``feature'' at the glass transition,  and that   the second derivative of the specific heat is the first derivative  akin to a divergence, albeit   with large finite-size effects. For this reason, we did  not try to identify the location of the phase transition using derivatives of the free energy.

\subsection{Coarse Estimate of Phase-Transition Temperature from Pinch Free Energies}

\begin{figure}
\includegraphics[width=\columnwidth]{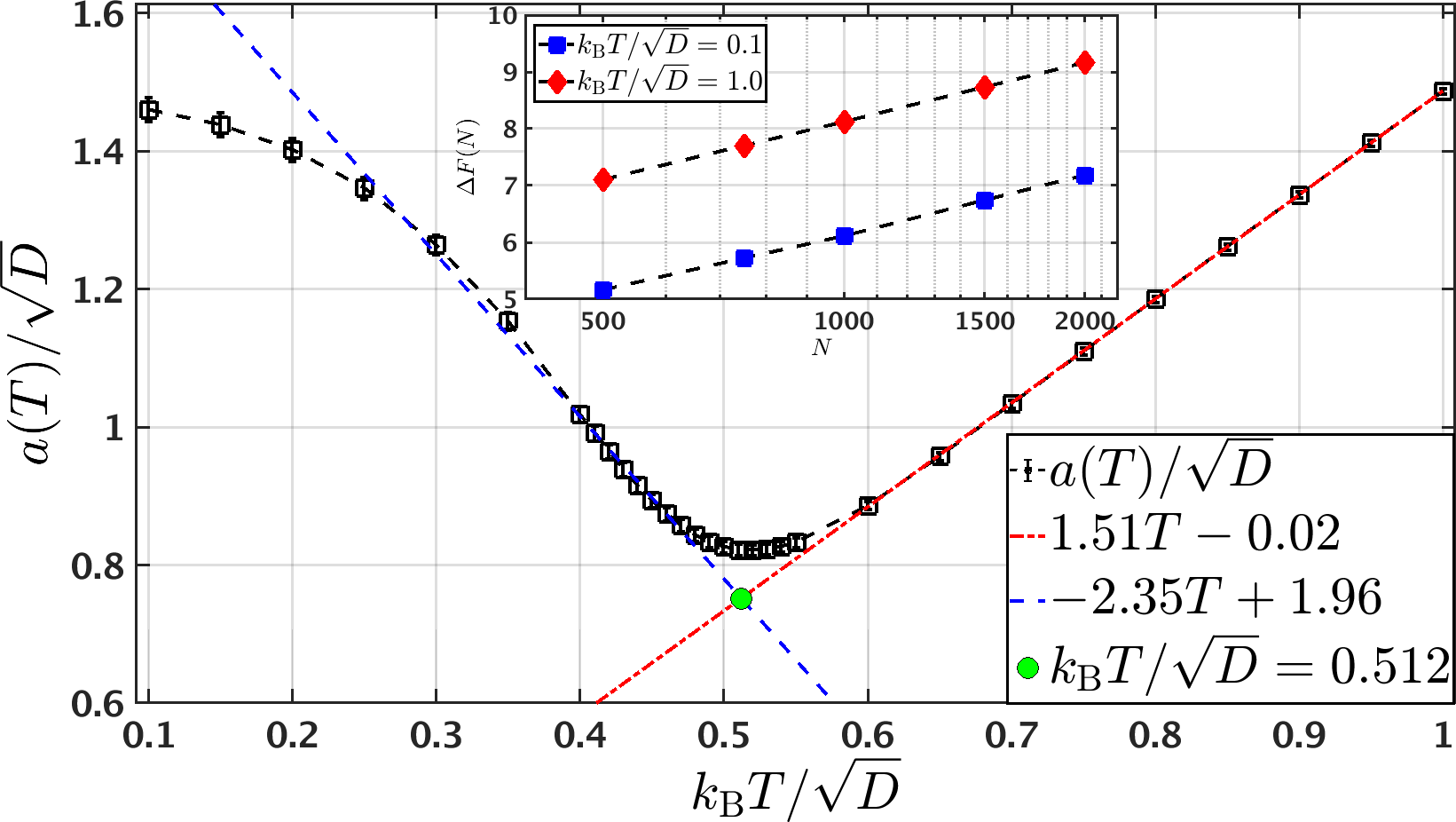}
\caption{\label{fig:pinchfreeenergy} Pinch free energy prefactor $a(T)$ versus temperature. The inset shows the disorder-averaged pinch free energy versus $\ln(N)$ in both the glass phase ($k_{\rm B}T=0.1\sqrt{D}$) and the molten phase ($k_{\rm B}T=1.0\sqrt{D}$). This allows us to extract the prefactors of these logarithmic behaviors, which are shown as a function of temperature in the main graph.  The temperature at which the prefactor changes slope is an estimate of the phase transition temperature $T_{\rm c}$.}
\end{figure}

To obtain a first estimate of the transition temperature, we follow  \cite{bund2002pre} and consider the disorder-averaged pinch free energy $\langle \Delta F \rangle$.  As seen in the inset of Fig.~\ref{fig:pinchfreeenergy}, this quantity has a logarithmic dependence on the sequence length $N$ at both low and high temperatures~\cite{bund2002pre}.  We thus fit it to the linear form
\begin{equation}
	\langle \Delta F(N) \rangle = a(T) \ln(N) + c(T).
\end{equation}
for sequence lengths in the range 500 $\leq N \leq$ 2000. The resulting prefactor $a(T)$ is shown in the main part of Fig.~\ref{fig:pinchfreeenergy}. As expected~\cite{pgg}, in the molten phase $a(T) \approx \frac{3}{2}k_{\rm B}T$ resulting in a linear dependence of the prefactor on temperature.  However, for lower temperatures, the prefactor is non-monotonic in temperature and thus lends itself as an   estimator of the critical temperature. The intersection of linear fits on either side of the minimum yields   $T_{\rm c} \approx 0.51\sqrt{D}/k_{\rm B}$. 

\subsection{Order Parameters for the Transition, and a More Precise Estimation of the Phase-Transition Temperature}
\begin{figure}[t]
{\includegraphics[width=\columnwidth]{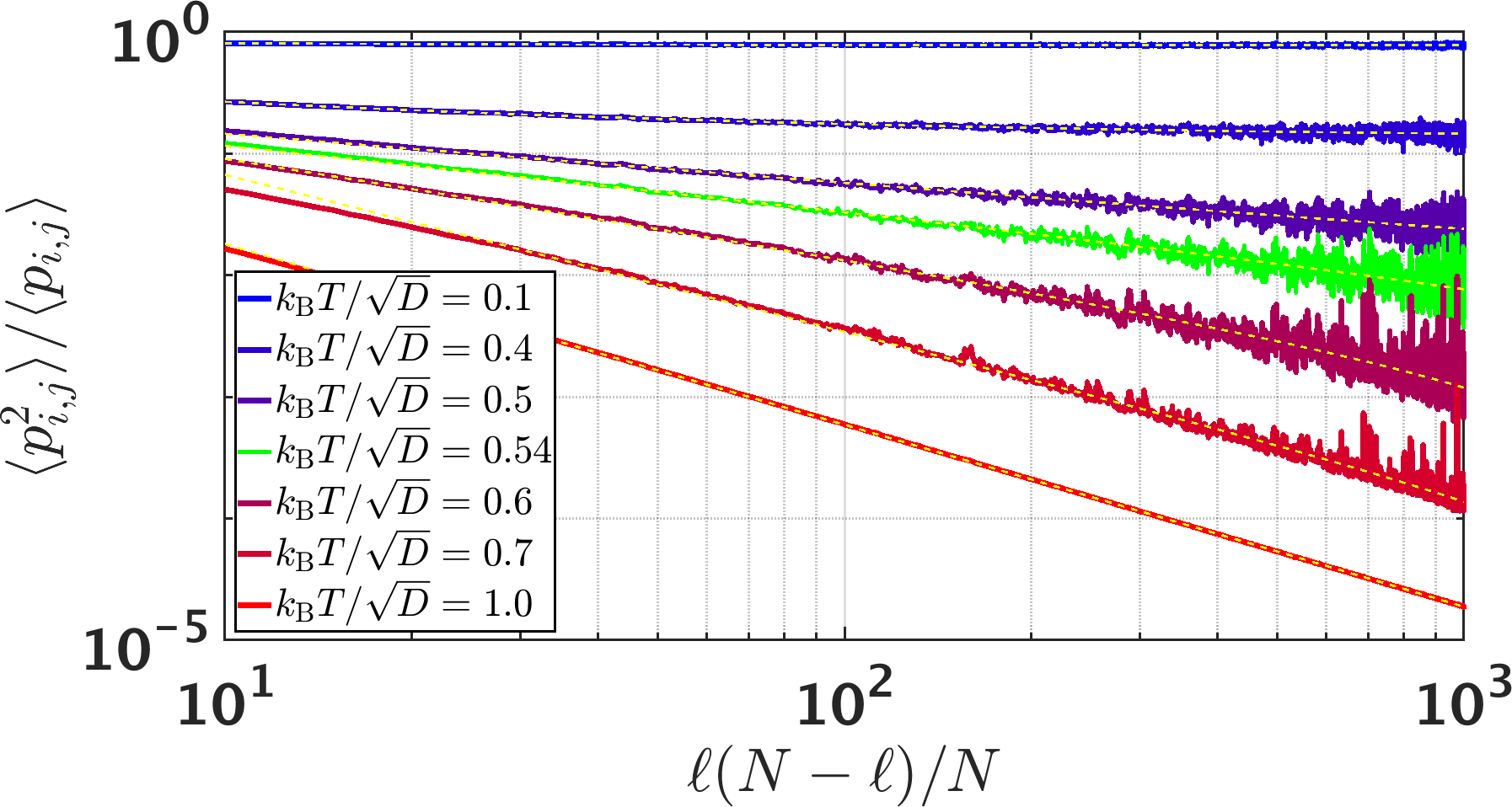}}
\caption{High-temperature pairing ratio $ {\left<p(\ell)^{2}\right>}/{\left<p(\ell)\right>}$ versus substrand length. The curves show the behavior of this new observable at several temperatures   in the vicinity of $T_{\rm c}$ and for temperatures deep within the glass phase ($k_{\rm B}T/\sqrt{D}=0.1$) and molten phase ($k_{\rm B}T/\sqrt{D}=1.0$).  The dashed yellow lines are fits to Eq.~(\protect\ref{eq:highTemp_ratio}).}
\label{fig:Psq_over_P_high-low_Ts}
\end{figure}

\begin{figure}[t]\vspace*{-.2mm}
{\includegraphics[width=\columnwidth]{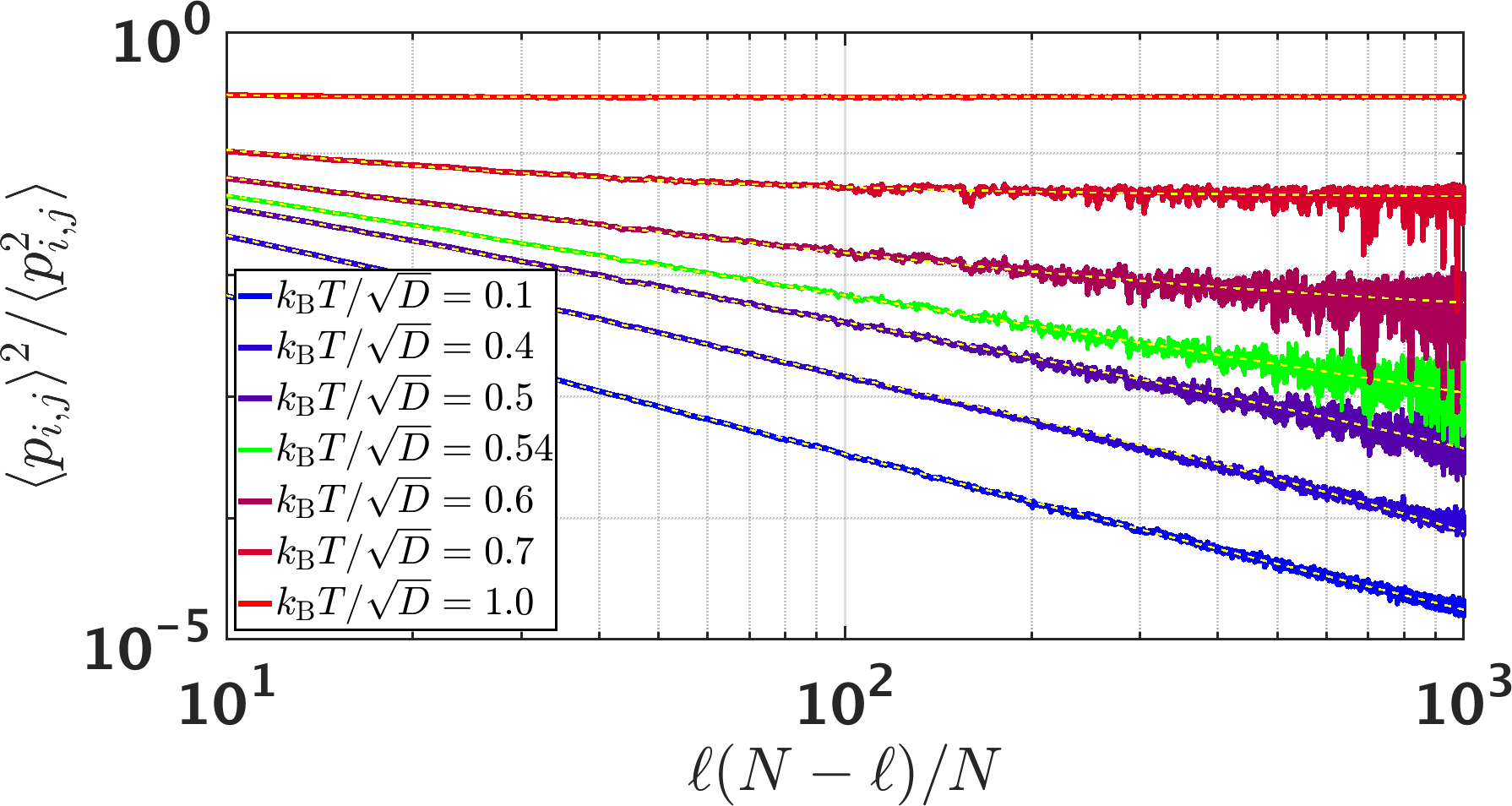}}

\vspace*{-1mm}
\caption{Low-temperature pairing ratio ${\left<p(\ell)\right>^{2}}/{\left<p(\ell)^{2}\right>}$ versus substrand length. Shown is the behavior of this new observable at several temperatures   in the vicinity of $T_{\rm c}$ and for temperatures deep within the glass phase ($k_{\rm B}T/\sqrt{D}=0.1$) and molten phase ($k_{\rm B}T/\sqrt{D}=1.0$). The dashed yellow lines are fits to Eq.~(\protect\ref{eq:lowTemp_ratio}).}
\label{fig:P-sq_over_Psq_high-low_Ts}
\end{figure}

Using this value of $T_{\rm c}$ as a guide, we limited the range over which we performed our simulations to $0.4 \leq k_{\rm B}T/\sqrt{D} \leq 0.7$ for sequence lengths up to $N=4000$. Since the scaling of the base-pairing probability $\langle p(\ell)^{n}\rangle$ by itself does not demonstrate a clear indication of when a finite system of length $N$ approaches $T_{\rm c}$, we defined two additional observables that  allow us to extract when the system approaches $T_{\rm c}$ from either side. The first is the ratio $\langle p(\ell)^{2} \rangle$/$\langle p(\ell)\rangle$, which we would expect to be constant for large $\ell$ in the glass phase due to $\alpha_2=\alpha_1$ and revert back to a power law similar to Eq.~(\ref{eq:pbindscaling}) with an exponent of ${3}/{2}$ in the molten phase.  Conversely, the second new observable, $\langle p(\ell) \rangle^{2}$/$\langle p(\ell)^{2}\rangle$ we  expect to be constant in the high-temperature regime for large $\ell$ and then decay with the power law of Eq.~(\ref{eq:pbindscaling}) and an exponent of approximately $1.36$ at low temperatures. Figs.~\ref{fig:Psq_over_P_high-low_Ts} and~\ref{fig:P-sq_over_Psq_high-low_Ts} show these observables for several temperatures, both above, below, and close to the critical temperature at $N=4000$. As indicated  with dashed yellow lines, we fit these curves to the forms 
\begin{eqnarray}\label{eq:highTemp_ratio}
\frac{\left<p(\ell)^{2}\right>}{\left<p(\ell)\right>} &=& A_g\left\lbrace\left[\frac{\ell(N-\ell)}{N}\right]^{-\omega^{(g)}}+\Delta_{g}\right\rbrace  \\
\label{eq:lowTemp_ratio}
\frac{\left<p(\ell)\right>^{2}}{\left<p(\ell)^{2}\right>} &=& A_m\left\lbrace\left[\frac{\ell(N-\ell)}{N}\right]^{-\omega^{(m)}}+\Delta_{m}\right\rbrace \ , \qquad
\end{eqnarray}
with offsets $\Delta_g$ and $\Delta_m$, exponents $\omega^{(g)}$ and $\omega^{(m)}$, and global prefactors $A_g$ and $A_m$. We identify the offsets $\Delta_{g}$ and $\Delta_{m}$ as the order parameters of the system. 

It may seem surprising to have {\em two} order parameters, one for each phase. It is not commonly known that such observables can  be constructed quite generally. In appendix~\ref{sec:ising}  we illustrate   this  fact for a two-dimensional Ising model, where it is analytically known that there is only one phase transition. As that example shows, an order parameter is a quantity that vanishes in one phase, is non-zero in the other one, and can be used to determine the critical temperature; it may, or may not, be associated with a specific symmetry.  The two quantities $\Delta_g$ and $\Delta_m$ we constructed each vanish in one of the phases and are non-zero in the other one, as they probe specific properties of their respective phases.  While we cannot exclude that there are two consecutive phase transitions at very close temperatures, we have no indication for this hypothesis. The example of the Ising model suggests that there is likely only one transition,  even though it is mathematically possible to have $\langle p(\ell)\rangle^2 < \langle p(\ell)^2\rangle < \langle p(\ell)\rangle$, which would make both quantities vanish simultaneously, and possibly for more than a single temperature.

\begin{figure*}
\includegraphics[width=2\columnwidth]{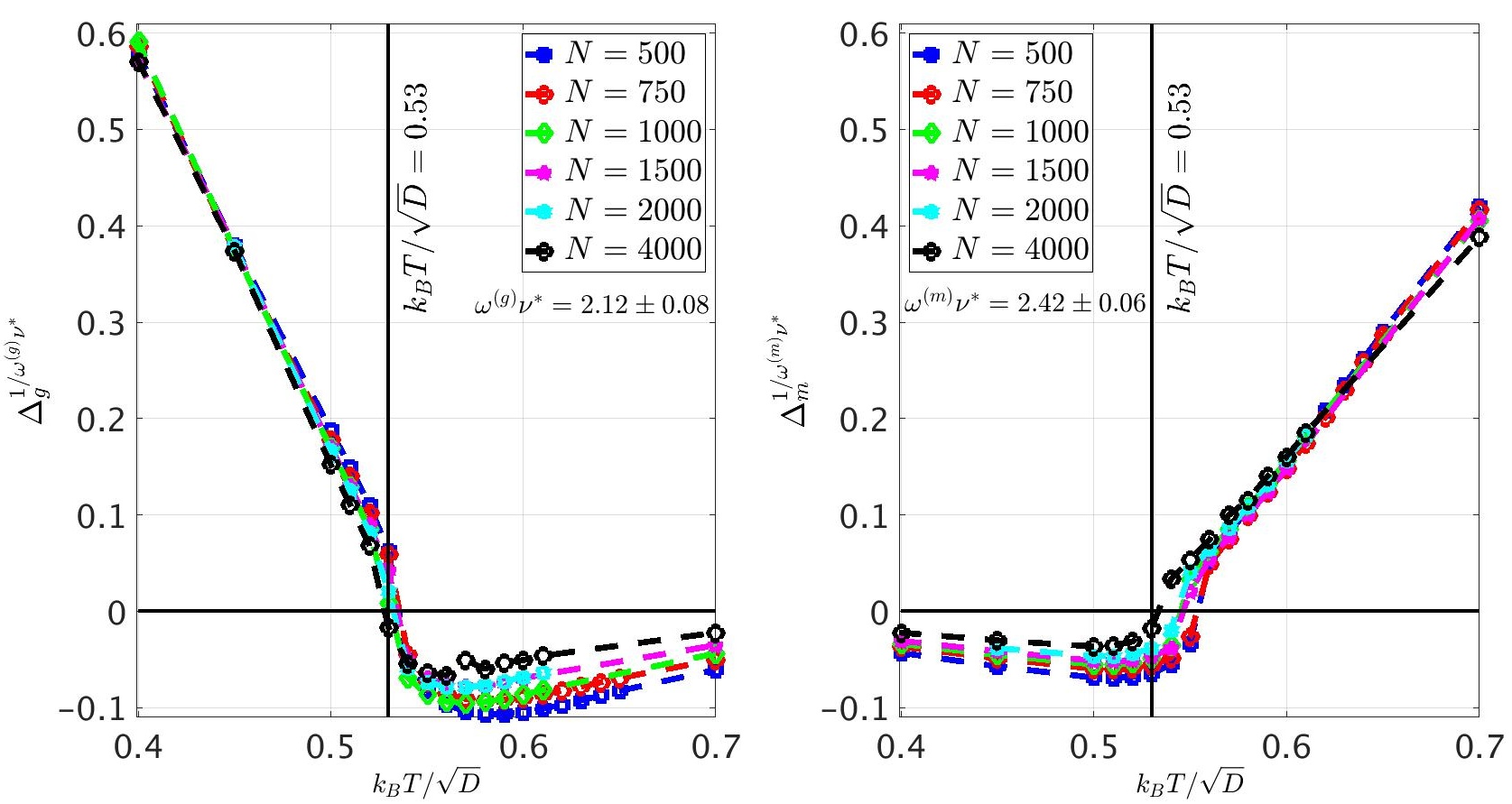}
\caption{\label{fig:Deltag_Deltam_for-all-Ns} Appropriate powers of the order parameters $\Delta_{g}$ and $\Delta_{m}$ versus temperature. (a) As $\Delta_g$ defined by Eq.~(\ref{eq:lowTemp_ratio}) approaches $k_{\rm B}T/\sqrt{D}\approx0.53$ from below it drops below zero. (b) Similarly, as $\Delta_m$ defined by Eq.~(\ref{eq:highTemp_ratio}) approaches $k_{\rm B}T/\sqrt{D}\approx0.53$ from above it becomes negative. Note that in order to take the appropriate powers of the order parameters, the power is taken of their absolute value and the sign is restored afterwards.  On the side of the transition, where the order parameters $\Delta_g$ and $\Delta_m$ are negative, they tend to zero with increasing system size $N$. This allows precise determination of the phase transition temperature $k_{\rm B}T/\sqrt{D}\approx0.53\pm0.01$.}
\end{figure*}

The order parameters $\Delta_{g}$ and $\Delta_{m}$ scale close to the transition as $\xi^{-\omega}$, where $\xi$ is a characteristic length scale, diverging at the transition as $\xi\sim |T-T_{\rm c}|^{-\nu^*}$. This implies that $\Delta_{g}$ and $\Delta_{m}$ should depend on temperature $T$ as 
\begin{subequations}
        \label{subeqn:Deltas}
        \begin{align}
                \Delta_{g} \sim \vert T-T_{\rm c}\vert^{\omega^{(g)}\nu^*} \label{subeqn:Delta_g},\\
                \Delta_{m} \sim \vert T-T_{\rm c}\vert^{\omega^{(m)}\nu^*} \label{subeqn:Delta_m}.
        \end{align}
\end{subequations}
To determine the two exponents,  we vary them  over a range of $1\leq(\omega^{(g)}\nu^*, \omega^{(m)}\nu^*)\leq4$ for each system size $N$ and fit a linear regression to $\Delta_{g}^{(\omega^{(g)}\nu^*)^{-1}}$ versus $T$, and $\Delta_{m}^{(\omega^{(m)}\nu^*)^{-1}}$ versus $T$ for a small temperature range immediately below and above $k_{\rm B}T/\sqrt{D}=0.53$, respectively. We then choose for each system size the values of $\omega^{(g)}\nu^*$ and $\omega^{(m)}\nu^*$ that maximize the coefficient of determination $R^2$ of the linear fits.  Lastly, we treat each system size as an independent realization (upon inspection the dependence of the optimal $\omega^{(g)}\nu^*$ and $\omega^{(m)}\nu^*$ on system size does not seem to have a systematic contribution) and determine the average and standard error of the mean of the two exponents. On the left side of the critical point ($T<T_{\rm c}$), we find
\begin{equation}\label{eq:omegagnunumerical}
\omega^{(g)}\nu^*\approx2.12\pm 0.08,
\end{equation}
while on its right side ($T>T_{\rm c}$)
\begin{equation}\label{eq:omegamnunumerical}
\omega^{(m)}\nu^*\approx2.43\pm 0.06.
\end{equation}
Fig.~\ref{fig:Deltag_Deltam_for-all-Ns} shows $\Delta_g^{\frac{1}{\omega^{(g)}\nu^*}}$ and $\Delta_m^{\frac{1}{\omega^{(m)}\nu^*}}$ with the such determined values for $\omega^{(g)}\nu^*$ and $\omega^{(m)}\nu^*$ as functions of temperature.  One clearly sees that each quantity approaches zero roughly linearly on its respective side of $k_{\rm B}T_{\rm c}/\sqrt{D}=0.53$ ($\Delta_g$ for $T<T_{\rm c}$ and $\Delta_m$ for $T>T_{\rm c}$).  Both cross zero at $k_{\rm B}T_{\rm c}/\sqrt{D}=0.53$ and become dependent on the system size $N$ on the opposite side of $k_{\rm B}T_{\rm c}/\sqrt{D}=0.53$, where they remain small and appear to tend toward zero with increasing system size $N$.  We thus conclude that $\Delta_g$ and $\Delta_m$ are non-zero on their respective side of the phase transition and vanish on the opposite side, rendering them proper order parameters for the transition.  We further conclude that the phase transition takes place at $k_{\rm B}T_{\rm c}/\sqrt{D}=0.53\pm0.01$.
In order to verify the consistency of the observed critical behavior, we note that the field theoretical predictions given in Eqs.~(\ref{eq:nuvalue}) and  (\ref{eq:moltenalphavalues})-(\ref{eq:glassalphavalues}) read
\begin{equation}
\omega^{(g)}\nu^*=(2\alpha_1^{(g)}-\alpha_2^{(g)})\nu^*=1.36\cdot1.56 = 2.12
\end{equation}
and
\begin{equation}
\omega^{(m)}\nu^*=(\alpha_2^{(m)}-\alpha_1^{(m)})\nu^*=\frac{3}{2}\cdot1.56 = 2.34\ .
\end{equation} 
These values are in good agreement with our numerically determined values in Eqs.~(\ref{eq:omegagnunumerical}) and~(\ref{eq:omegamnunumerical}).

\section{Transition Mechanism} 
\label{sec:mechanism}

With the critical temperature in hand, we aim to understand the mechanism of the phase transition itself. Deep in the molten phase, base-pairing energies are irrelevant and as a consequence any base can pair with any other base. Thus, the thermally averaged base-pairing probability $p(\ell)$ is the same for any pair of bases with the same distance $\ell$ and decays with the distance $\ell$ as the power law introduced in Eq.~(\ref{eq:pbindscaling}). This probability is independent of the disorder realization.  In contrast, at zero temperature the RNA molecule folds into the ground-state structure, which is typically non-degenerate within the Gaussian-disorder model.  Thus, for a given realization of the disorder the thermally averaged   pairing probability of a given base pair is either zero (if the base pair is not in the ground-state structure) or one (if the base pair is in the ground-state structure). Which base pairs have a zero probability and which have a probability of one changes with the disorder realization. Thus, the scaling behavior  (\ref{eq:pbindscaling}) at zero temperature arises only {\em after} ensemble averaging.

As discussed in Sec.~\ref{subsec:lwreview}, and illustrated in Fig.~\ref{fig:locking_figure}, the picture of the phase transition proposed in~\cite{mlkw} is that as the phase transition is approached from below, the fraction of base pairs ``locked'' into a ground-state structure decreases by the appearance of molten regions. At the transition, the situation inverts and the ``locked'' base pairs become localized regions of decreasing size within an overall molten structure on the high-temperature side of the phase transition. These ideas are reflected in the order parameters introduced in Eqs.~(\ref{eq:highTemp_ratio}) and (\ref{eq:lowTemp_ratio}). Let us stress that while the expression ``locked'' used in Ref.~\cite{mlkw} suggests that bases are paired with probability $1$,  this condition can be relaxed to mean that they are paired with probability $p>0.5$, or even $p\ge 0.1$. This suffices to render the expectations (\ref{eq:highTemp_ratio}) and (\ref{eq:lowTemp_ratio}) non-trivial. We will understand ``locked'' in this sense from now on.

\medskip

\subsection{Base-pairing probability distribution}

In order to directly probe the transition mechanism, we numerically calculate the entire {\em distribution} of base-pairing probabilities $P(p;\ell)$.  We note that all moments of the disorder-averaged base-pairing probabilities can be reconstructed from these distributions as
\begin{equation}
\langle p(\ell)^n \rangle = \int_{0}^{1} p^n P(p;\ell)\, \mathrm{d}p.
\end{equation}
We obtain these distributions by explicitly calculating the base-pairing probabilities $p_{i,i+\ell}$ using Eq.~(\ref{eq:pij}) and then tabulating their frequencies averaged over all $i$ and  many realizations of the disorder.  Moreover, due to the large   range of   base-pairing probabilities, instead of taking histograms of the $p_{i,i+\ell}$ themselves, we   sample the distribution $Q(x;\ell)$, where $x=-\ln(p(\ell))$.  The two distributions are connected to each other by $Q(x;\ell)\mathrm{d} x = P(p;\ell)\mathrm{d} p$
and thus moments of the disorder-averaged base-pairing probability can be obtained from $Q(x;\ell)$ as
\begin{equation}
\langle p(\ell)^n \rangle = \int_{0}^{\infty} e^{-nx} Q(x;\ell) \,\mathrm{d}x.
\end{equation}
To reduce the statistical noise, we further use 
\begin{equation}
Q(x):= \int_{N/3}^{N/2}Q(x;\ell) \,\mathrm{d} \ell,
\end{equation}
where the range of integration is motivated by the fact that according to Eq.~(\ref{eq:pbindscaling}), the pairing probability $p(\ell)$ is a function of $\ell(N-\ell)/N$, and in the chosen range the latter does not vary by more than $10 \%$.

\begin{figure}
\includegraphics[width=1\columnwidth]{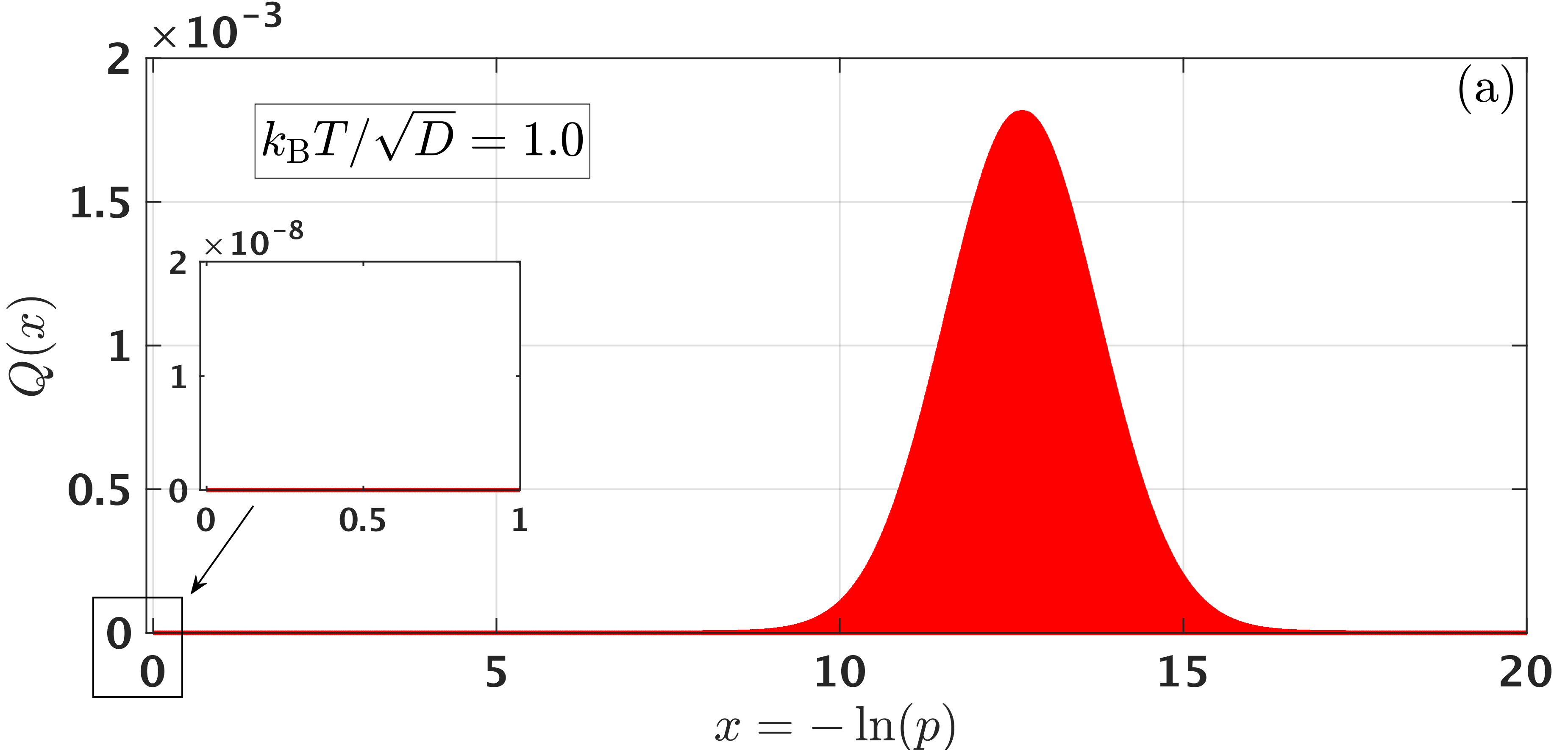}\\
\includegraphics[width=1\columnwidth]{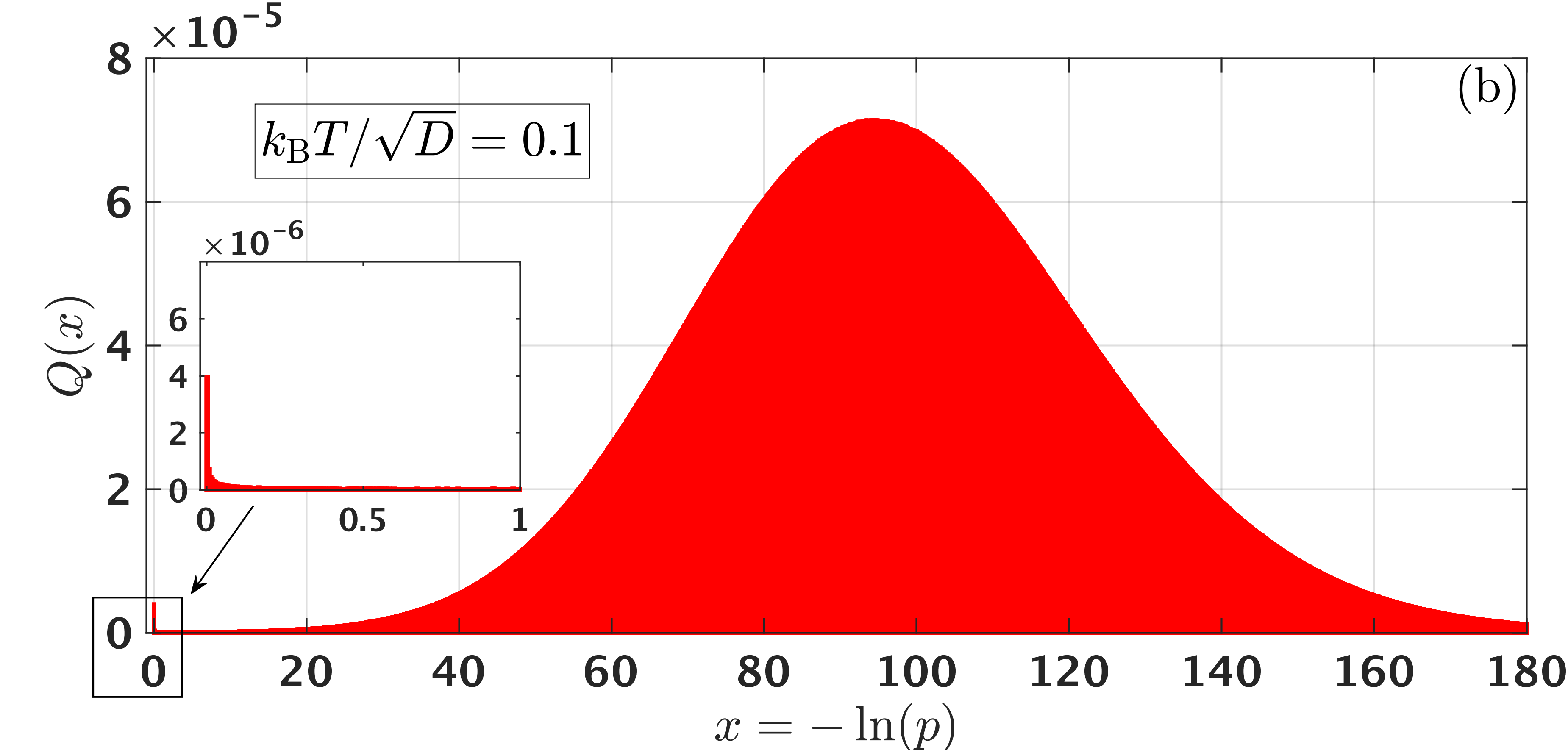}
\caption{Distributions of the logarithm of base-pairing probabilities $x=-\ln(p)$ for system size $N=4000$ averaged over the final third of all $\ell$'s. (a)  Deep in the molten phase ($k_{\rm B}T/\sqrt{D}=1.0$) the distribution of base-pairing probabilities has a distinct Gaussian-like form, which would go to a delta peak as $T\rightarrow\infty$ ($D\rightarrow 0$). The inset shows that no weight of the distribution is present close to $x\approx0$ ($p\approx1$). (b) Deep in the glass phase ($k_{\rm B}T/\sqrt{D}=0.1$) the base-pairing distribution still has a Gaussian-like form, albeit broader than the high-temperature distribution. As the inset shows there exists now a peak close to $x\approx0$ ($p\approx1$).}
\label{fig:avgQdisthighandlow}
\end{figure}

Fig.~\ref{fig:avgQdisthighandlow} shows that in both temperature regimes the base-pairing probability distribution $Q(x)$ has a Gaussian-like shape. Fig.~\ref{fig:avgQdisthighandlow}(a) is the distribution of $Q(x)$ at the highest temperature considered, $k_{\rm B}T/\sqrt{D}=1$, for a system of size $N=4000$. The distribution is centered around $x\approx13$ and we interpret it as a broadened version of what would be a $\delta$-peak around the value of $\langle p(\ell)\rangle$ from Eq.~(\ref{eq:pbindscaling}) at infinite temperature or $D=0$. Fig.~\ref{fig:avgQdisthighandlow}(b) shows the same distribution at the lowest temperature numerically accessible to us, $k_{\rm B}T/\sqrt{D}=0.1$. Here we see a broad Gaussian-like distribution again.  However, it is by far wider than the high-temperature distribution and is located at $x\approx95$, i.e., at much smaller base-pairing probabilities than its high-temperature counterpart.  We interpret this Gaussian-like feature as the contributions from all the base pairs that are {\em not} part of the ground-state structure and thus would have exactly zero probability at zero temperature, while acquiring a finite, albeit very small ($x\gg0$, $p\ll1$), pairing probability at low  temperatures.  Importantly, as the inset of Fig.~\ref{fig:avgQdisthighandlow}(b) shows, close to the origin ($x\approx0$ or $p\approx1$) there exists another peak. We interpret this prominent peak at $x\approx0$ as the contributions from those base pairs that are ``locked'' in the ground-state structure and thus define the glass phase of this model.  As the inset in Fig.~\ref{fig:avgQdisthighandlow}(a) shows, no corresponding peak at $x\approx0$ occurs deep in the molten phase.

\begin{figure}
\includegraphics[width=1\columnwidth]{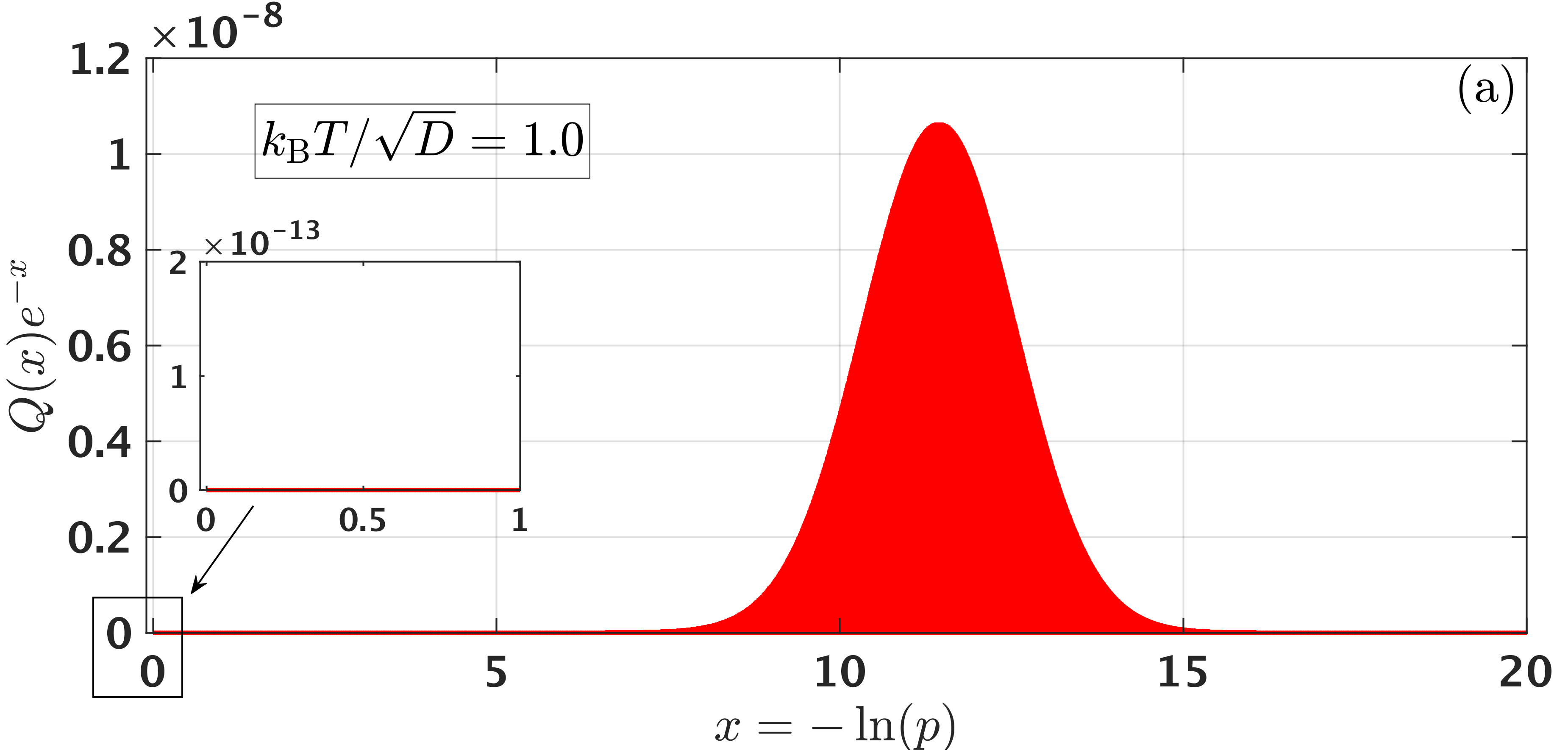}\\
\includegraphics[width=1\columnwidth]{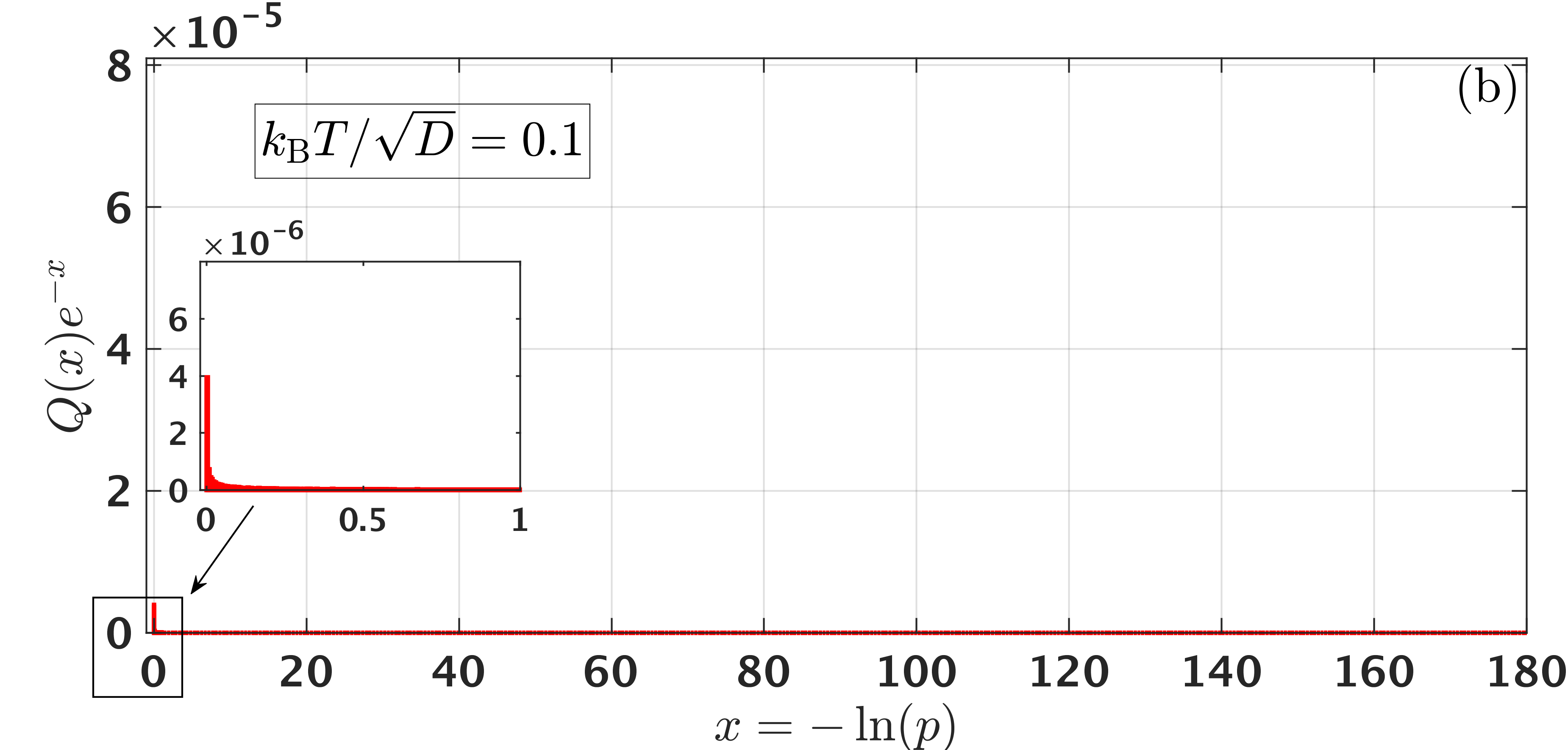}
\caption{Distributions of the logarithm of base-pairing probabilities $x=-\ln(p)$ for system size $N=4000$ averaged over the final third of all $\ell$'s multiplied with $e^{-x}$ for the same ranges as shown in Fig.~\protect\ref{fig:avgQdisthighandlow}, there without the factor of \protect$e^{-x}$. (a)  Deep in the molten phase ($k_{\rm B}T/\sqrt{D}=1.0$) the distribution of base-pairing probabilities has a distinct Gaussian-like form, which would go to a delta peak as $T\rightarrow\infty$ ($D\rightarrow 0$). (b) Deep in the glass phase ($k_{\rm B}T/\sqrt{D}=0.1$) the base-pairing distribution consists of only a peak at $x\approx0$ ($p\approx1$). Note the difference in  scales between figures \ref{fig:avgQdisthighandlow}(a) and \ref{fig:expQdisthighandlow}(a) while the scales in figures \ref{fig:avgQdisthighandlow}(b) and \ref{fig:expQdisthighandlow}(b) are identical.}
\label{fig:expQdisthighandlow}
\end{figure}

As discussed above, when calculating the ensemble averaged pairing probability $\langle p(\ell)\rangle$ the distribution $Q(x;\ell)$ has to be integrated against $e^{-x}$.  Thus, in order to appreciate where the weight contributing to the ensemble-averaged pairing probability $\langle p(\ell)\rangle$ is coming from in the two phases, Fig.~\ref{fig:expQdisthighandlow} shows $e^{-x}Q(x)$.  Deep in the molten phase ($k_{\rm B}T/\sqrt{D}=1.0$) multiplying with $e^{-x}$ simply shifts the peak of the Gaussian-like distribution to somewhat smaller $x$ (larger $p$).  In contrast, deep in the glass phase ($k_{\rm B}T/\sqrt{D}=0.1$) the Gaussian peak at large $x$ observed in the distribution $Q(x)$ alone becomes invisible and only the feature at $x\approx0$ remains and is unaffected in amplitude by multiplication with $e^{-x}$.  Thus, as suspected, in the glass phase only the small-$x$ range contributes significantly to the ensemble-averaged pairing probability.

\begin{figure}
\includegraphics[width=1\columnwidth]{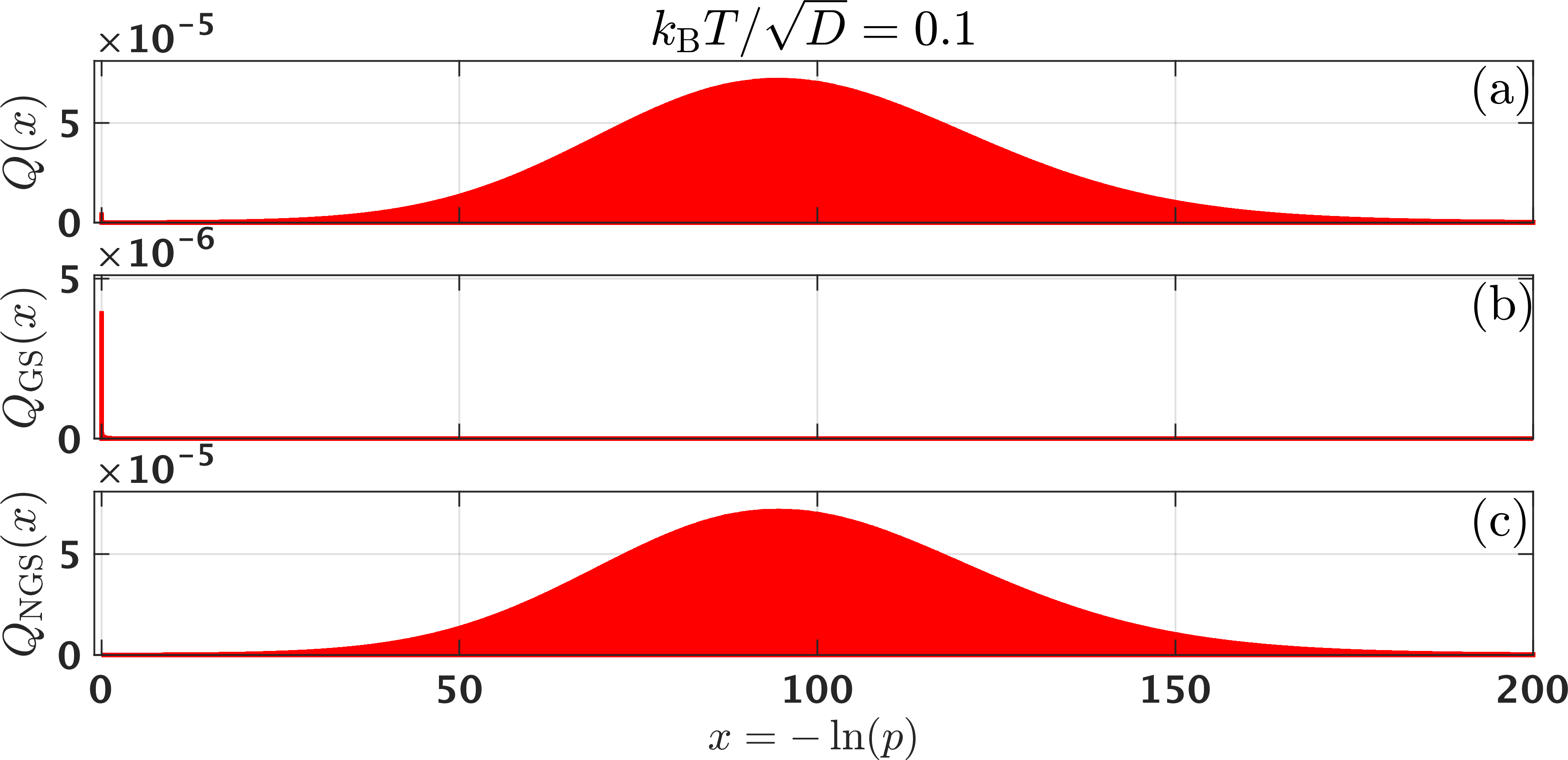}
\caption{Separation of the glass-phase ground and non-ground-state distributions. (a) The total base-pairing distribution within the glass phase ($k_{\rm B}T/\sqrt{D}=0.1$). (b) The ground-state base-pairing distribution (note the difference in scale). This peak represents the frequency of those pairs ``locked'' in the ground-state structure and are the contributing weight seen in the inset of Fig.~\ref{fig:avgQdisthighandlow}(b). (c) The non-ground-state base-pairing distribution. As mentioned in the text, the broad Gaussian-like peak seen in the low-temperature regime is a result of non-ground-state pairs that have a small, non-zero pairing probability at finite temperatures. These probabilities would be zero at $T=0$. \label{fig:avg_Tot_GS_NGS_lowT}}
\end{figure}

\begin{figure}
\includegraphics[width=1\columnwidth]{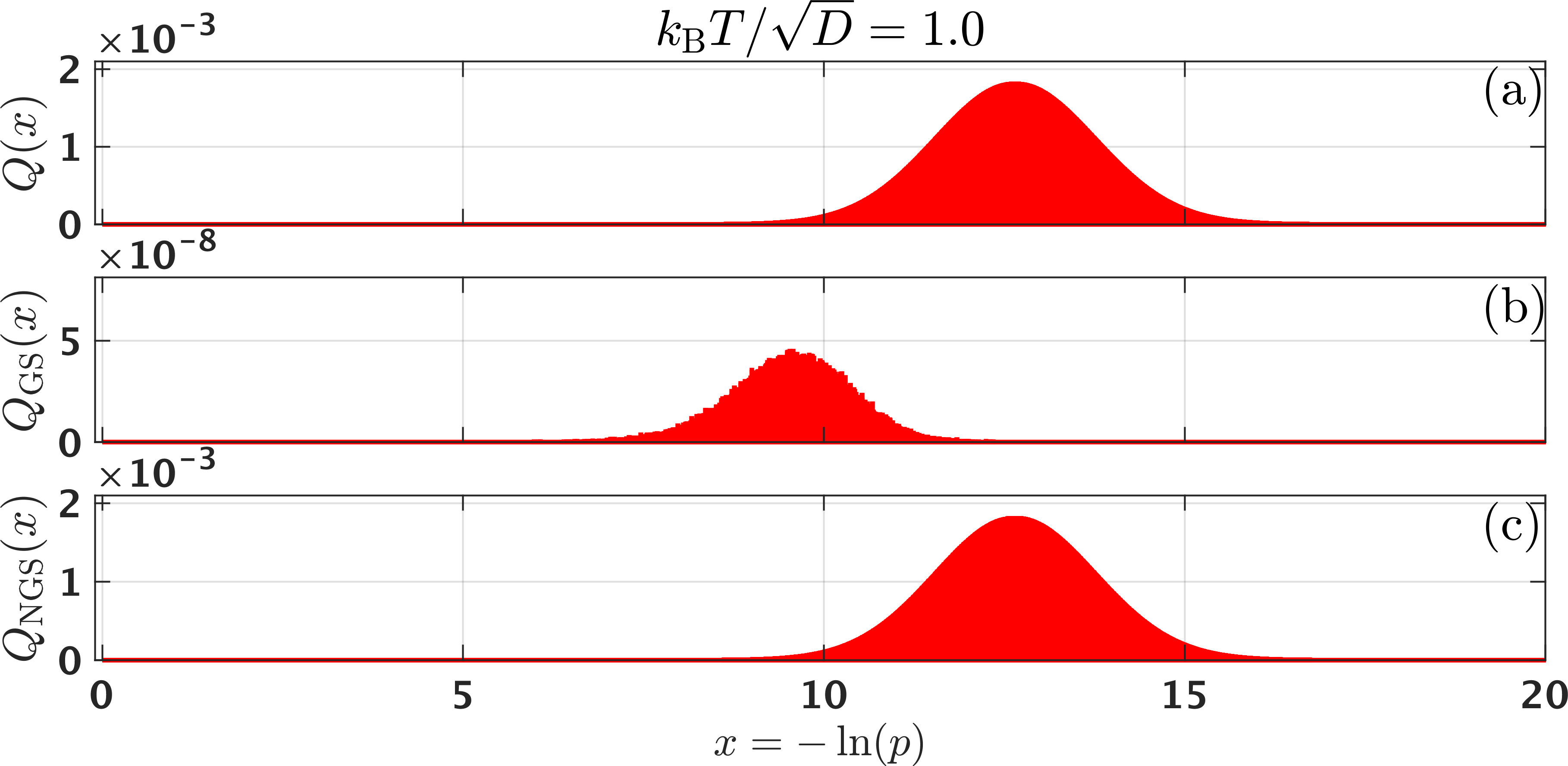}
\caption{Separation of the molten-phase ground and non-ground-state distributions. (a) The total base-pairing distribution within the molten phase ($k_{\rm B}T/\sqrt{D}=1.0$). (b) The ground-state base-pairing distribution (note the difference in scale). The lack of weight close to $x\approx0$ is due to a significantly reduced pairing probability of ground-state base pairs at high temperatures leading to the absence of any feature in the inset of Fig.~\ref{fig:avgQdisthighandlow}(a). (c) The non-ground-state pairs dominate the total pairing probability distribution. \label{fig:avg_Tot_GS_NGS_highT}}
\end{figure}

\subsection{Locked base pairs at finite temperature stem from ground-state base pairs at zero temperature}
In order to confirm that the peak at $x\approx0$ in the low-temperature behavior is indeed generated from the base pairs that form the ground-state structure at $T=0$, we explicitly separate the distribution $Q(x)$ into two sub-distributions. We use the approach outlined at the end of Sec.~\ref{subsec:partfunc} to calculate the ground-state structure for every disorder configuration and then tabulate the base-pairing probabilities across all $\ell$ for base pairs participating in the ground-state structure separately from those which do not. This allows us to split the entire base-pairing probability $Q(x)$ into a ground-state contribution $Q_{\rm GS}(x)$ and a non-ground-state contribution $Q_{\rm NGS}(x)$. Figs.~\ref{fig:avg_Tot_GS_NGS_lowT} and~\ref{fig:avg_Tot_GS_NGS_highT} show these ground-state and non-ground-state distributions at the lowest and highest temperatures considered. In Fig.~\ref{fig:avg_Tot_GS_NGS_lowT}(b) one can see that the ground-state pairs directly contribute to the distinct peak close to $x\approx0$, which we interpret as   those base pairs that are ``locked'' in the ground-state structures. This further confirms that it is the non-ground-state pairs that make up the bulk of the total distribution seen in Fig.~\ref{fig:avgQdisthighandlow}(b), each of which would otherwise be zero at $T=0$. Fig.~\ref{fig:avg_Tot_GS_NGS_highT} then explains the absence of any weight in the inset of Fig.~\ref{fig:avgQdisthighandlow}(a). The lack of ground-state pair contributions in the region $x\approx0$ indicates that the system is no longer in the glass phase at this temperature.  Since the transition is driven by the behavior of the ground-state base pairs alone, we  focus on $Q_{\rm GS}(x)$ from now on.

\begin{figure}
\includegraphics[width=1\columnwidth]{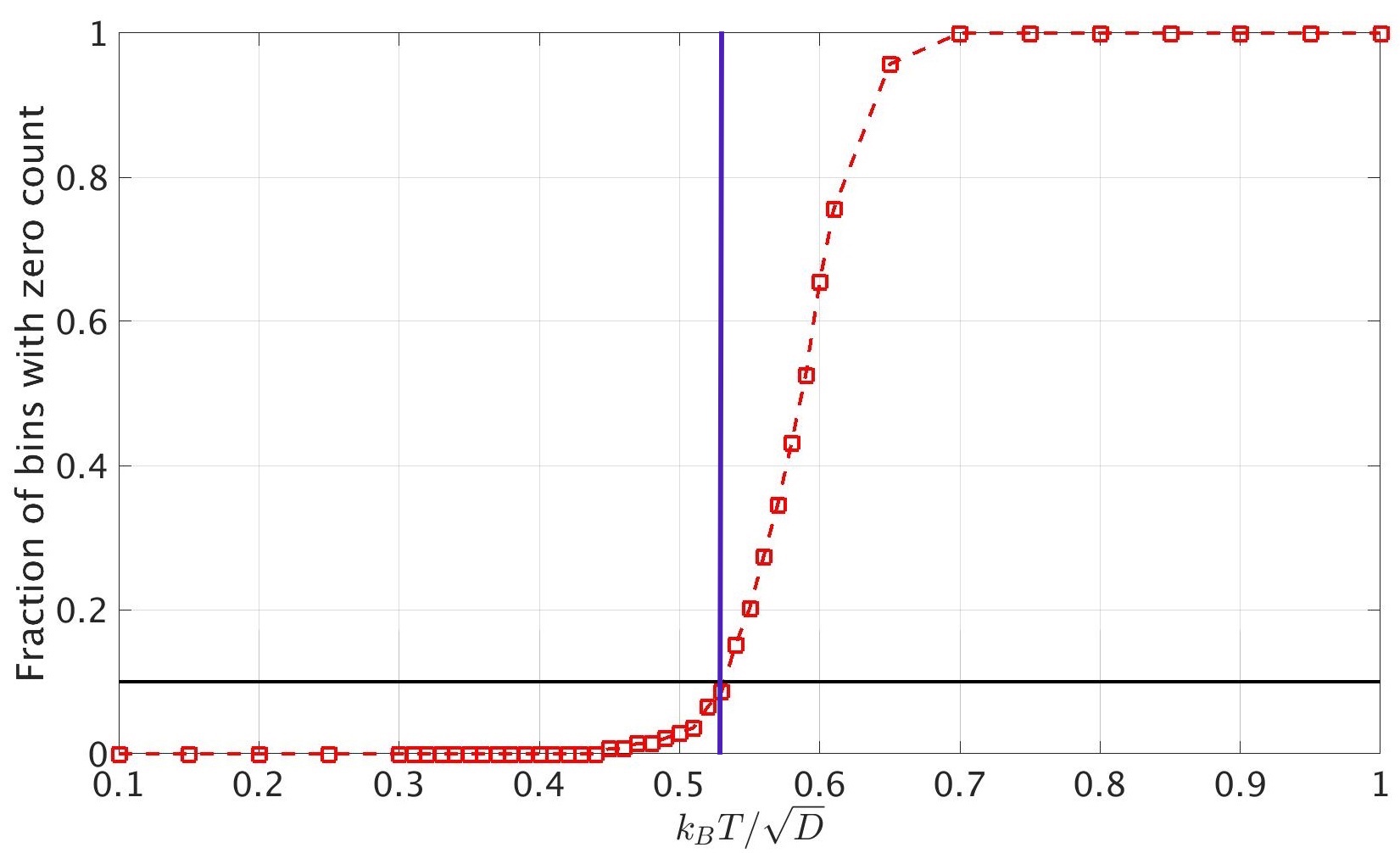}
\caption{Fraction of bins at $x\le0.7$ with zero counts as a function of temperature for $N=1000$. We split the range $0\le x < 0.7$ (corresponding to $p\ge0.5$) into $139$ equal bins of width $0.005$. A bin has zero counts when not a single ground-state base pair with $N/3\le\ell\le N/2$ in $10{,}000$ disorder configurations has a thermally averaged base pairing probability in the interval of $x$ associated with the bin. The fraction of bins with zero counts dramatically increases from near zero to near one at the phase transition temperature.  The horizontal solid line indicates where 10\% of the bins have zero counts, the vertical line the location of $T_{\rm c}$.\label{fig:zerocountfraction}}
\end{figure}

\subsection{Locked base pairs disappear at the phase transition}\label{subsec:lockedatTc}
Our next goal is to quantify how the weight of ``locked'' ground-state base pairs changes through the transition. Since a ground-state base pair remaining ``locked'' at a finite temperature does not mean that the pairing probability of this base pair is one, but rather that it is sizable in some way, we introduce a pragmatic way to characterize the disappearance of ``locked'' base pairs. Specifically, we divide the range $0\le x \le0.7$ (corresponding to $p\ge0.5$) into an equal number of bins and ask which fraction of these bins receive at least one ground-state base pair with $N/3\le\ell\le N/2$ when we sample $10{,}000$ independent disorder configurations (for $N=1{,}000$). This quantity is shown in Fig.~\ref{fig:zerocountfraction}. While its detailed behavior  is definitely dependent on the choice of the upper cutoff on $x$, the width of the bins, and the number of  samples, it is quite clear from the very strong temperature dependence shown in Fig.~\ref{fig:zerocountfraction} that essentially all log-probabilities $x\le0.7$ are present for $k_{\rm B}T/\sqrt{D} \le 0.45$, while they are absent for $k_{\rm B}T/\sqrt{D} \ge 0.7$. This switch should only moderately depend on our particular choices. It  is consistent with the disappearance of locked base pairs at the phase transition.

\subsection{Behavior of the base-pairing probability distribution at $x\approx0$}

\begin{figure}
\includegraphics[width=\columnwidth]{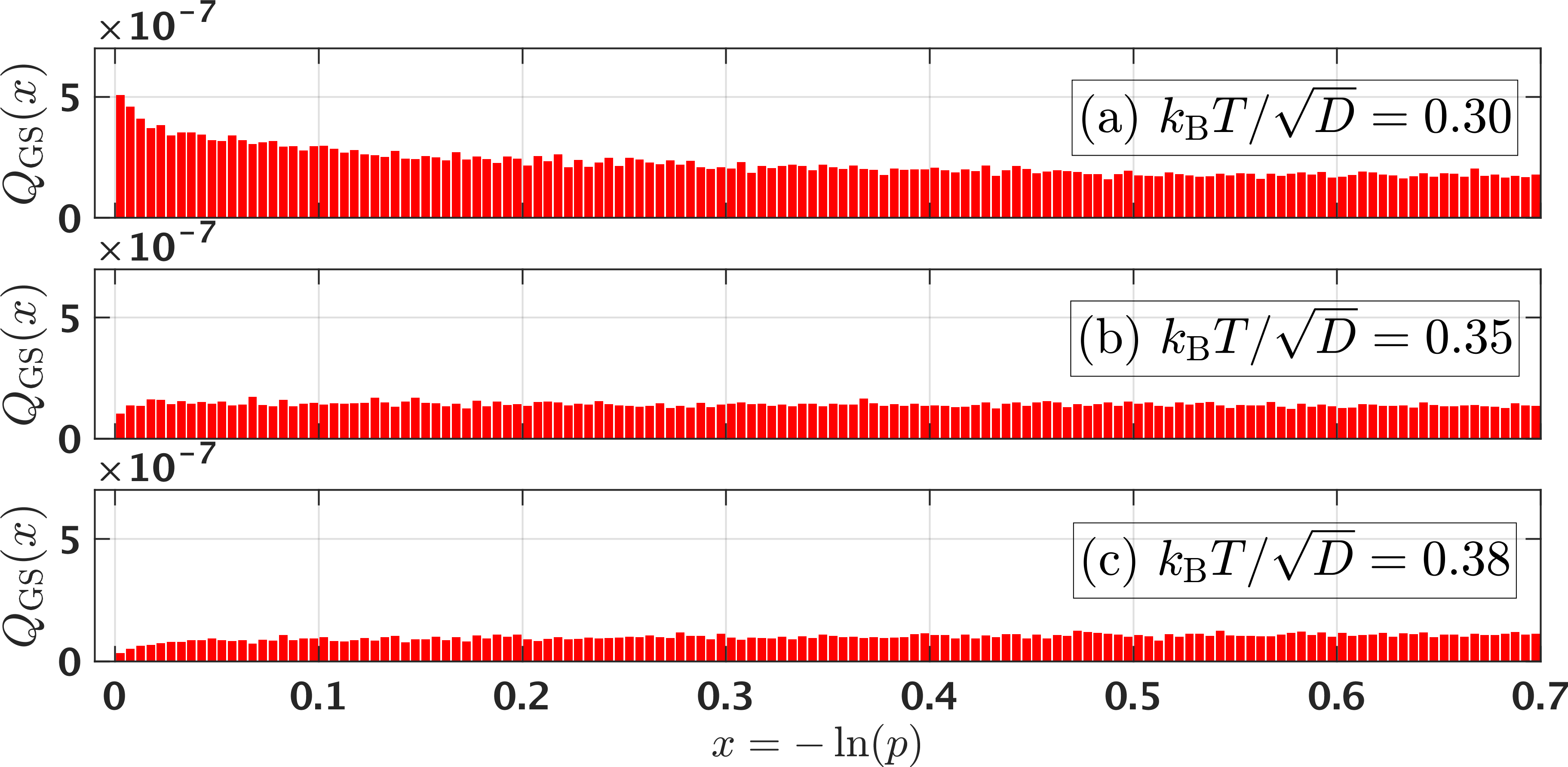}
\caption{\label{fig:avgQ_vs_x_N=1000_lowTemp} Average ground-state base-pairing probabilities versus $x$ for system size $N=1000$. (a) At $k_{\rm B}T/\sqrt{D}=0.3$, and all temperatures below, the region close to $x\approx0$ is the dominant peak. (b) As the temperature is increased, the height of these peaks all drop to roughly the same height as the neighboring peaks. (c) Above $k_{\rm B}T/\sqrt{D}\approx0.35$, the probability density $Q_{\mathrm{GS}}(x)$ at $x\approx0$ is actually lower than at moderate $x$.}
\end{figure}

\begin{figure}[b]
\includegraphics[width=\columnwidth]{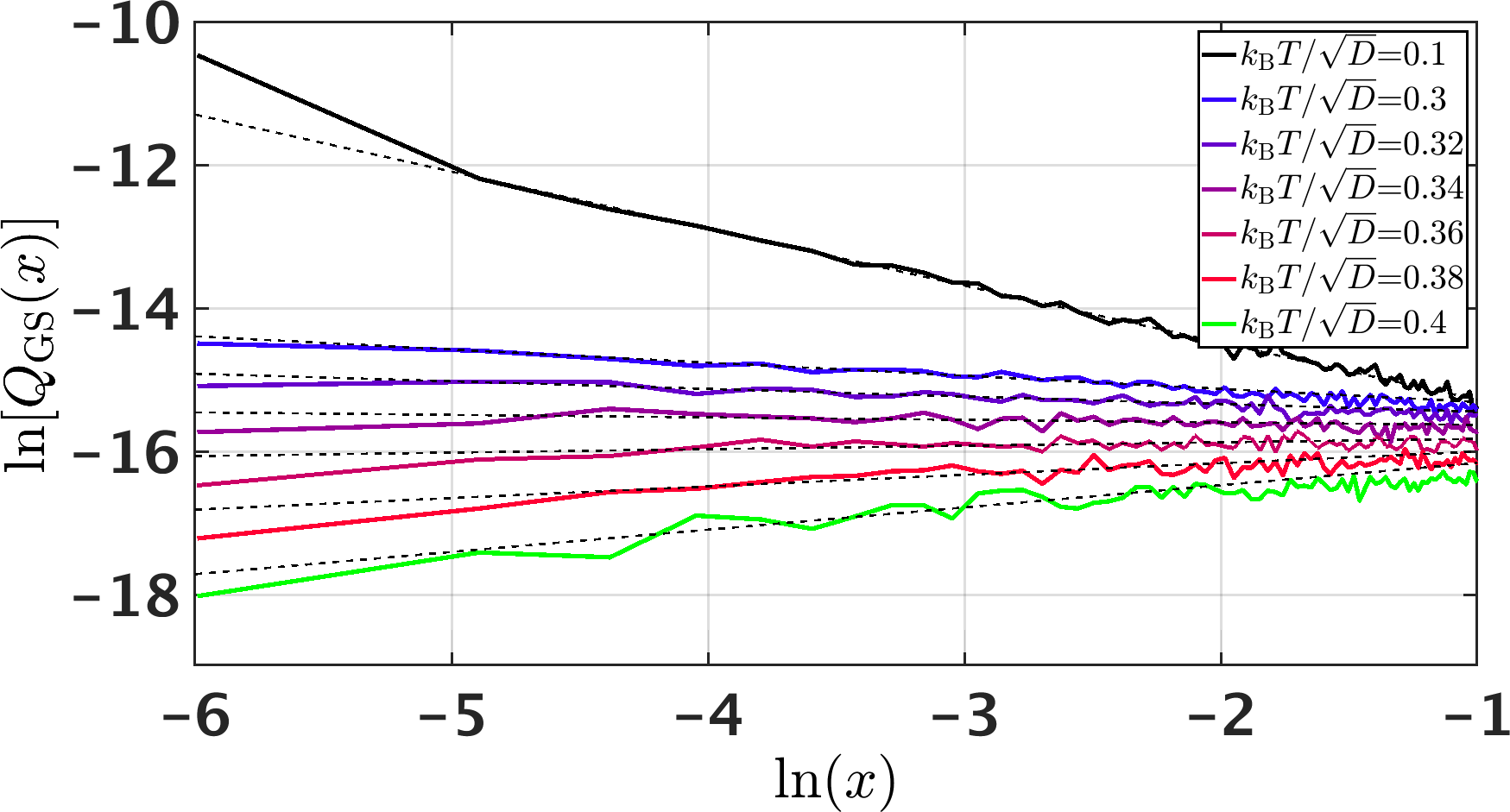}
\caption{\label{fig:logQ_vs_logx_N=1000_lowTemp} Double logarithmic plot of the average ground base-pairing distributions versus $x$ for system size $N=1,000$. At temperatures below the glass transition, the ground-state pairing distributions at moderate pairing probabilities are approximate power laws (black dashed fit lines, bins with zero counts ignored in the fits) whose exponents turn from negative to positive as the temperature increases.}
\end{figure}

As we have seen in Fig.~\ref{fig:avg_Tot_GS_NGS_lowT}, the base-pairing probability distribution is highly peaked at $x\approx0$ deep in the glass phase, consistent with a picture in which a base pair that is part of the ground-state structure at zero temperature still has probability essentially $1$ at $k_{\rm B}T/\sqrt{D}=0.1$.  Since we have seen in Sec.~\ref{subsec:lockedatTc} that the entire base-pairing probability  at $x\le0.7$ ($p\ge0.5$) disappears at the transition to the molten phase, it is of interest to look at the shape of the distribution as the transition is approached. Fig.~\ref{fig:avgQ_vs_x_N=1000_lowTemp} shows these distributions at various temperatures below, but close to, the glass-transition temperature. The most striking feature is that they switch from a behavior where the bulk of the distribution is at $x\approx0$ ($p=1$) at low temperatures to a behavior where base-pairing probabilities with small $x$ are still prevalent but probabilities at $x\approx0$ ($p\approx1$) itself are suppressed. An intermediate case of a distribution that is essentially independent of $x$ over the entire range $0\le x\le 0.7$ appears at $k_{\rm B}T/\sqrt{D}\approx0.35$.

\begin{figure}
\includegraphics[width=\columnwidth]{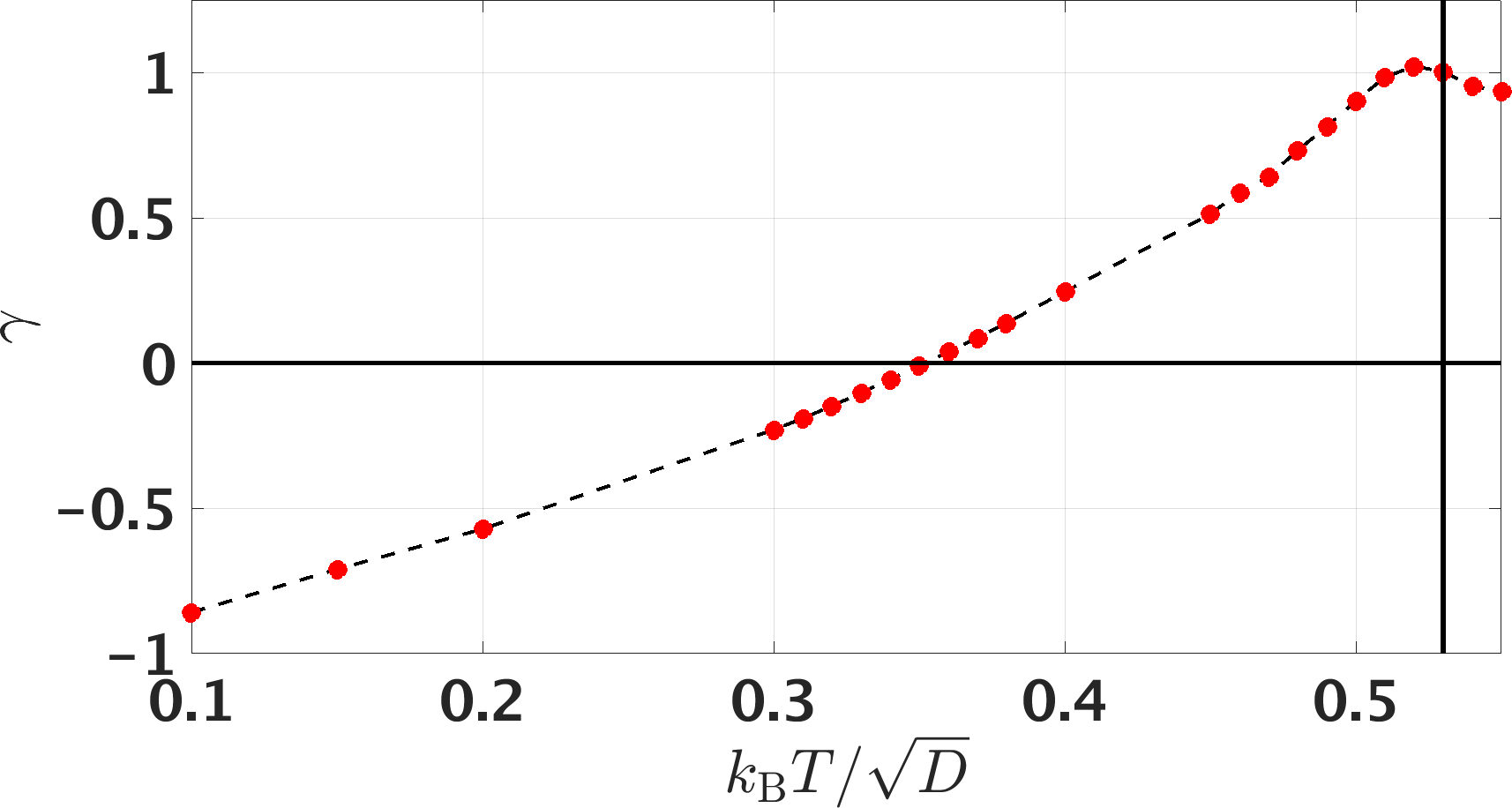}
\caption{\label{fig:gamma_vs_Temp_N=1000} $\gamma$ versus temperature. When plotted against temperature, the slopes of the fit lines from Fig.~\ref{fig:logQ_vs_logx_N=1000_lowTemp} cross zero at a point between $k_{\rm B}T=0.34\sqrt{D}$ and $k_{\rm B}T=0.36\sqrt{D}$, which   could be interpreted as the {\em secondary freezing} event mentioned in~\cite{mlkw}.}
\end{figure}

When plotting the same data on a double logarithmic scale (Fig.~\ref{fig:logQ_vs_logx_N=1000_lowTemp}) we find that over the range of $0\le x\le 0.7$ the ground-state base-pairing distribution is well described by a power law
\begin{equation}
Q_{\rm GS}(x)\sim x^{-\gamma}\ ,
\end{equation}
that allows us to capture its behavior in terms of a single temperature-dependent exponent $\gamma$. Figure~\ref{fig:logQ_vs_logx_N=1000_lowTemp} shows that the slope of these fits (the black dotted lines) turns from negative to positive with increasing temperature. The temperature dependence of the exponent $\gamma$ is shown in Fig.~\ref{fig:gamma_vs_Temp_N=1000}. For $T\to0$ it approaches $-1$ consistent with a true $\delta$-peak of the distribution at base-pairing probability one.  The exponent crosses zero for $k_{\rm B}T/\sqrt{D}$ between $0.34$ and $0.36$ and tends to one at the glass-transition temperature, where it becomes impossible to determine it, since too many bins in $x$   receive zero counts (see above).  We conclude that while the transition, at which high-probability base pairs disappear is at $k_{\rm B}T/\sqrt{D}\approx0.53$ as shown in Sec.~\ref{sec:tcest}, already at a lower temperature of $k_{\rm B}T/\sqrt{D}\approx0.35$ the system transitions from an ensemble where the bulk of the pairing probability distribution is at $p=1$ to an ensemble where the probability distribution has a power law with positive exponent at $x=0$, i.e., where the probability density goes to zero for pairing probabilities of $p=1$.
We call this transition, close in spirit to the one proposed in \cite{mlkw}, {\em secondary freezing}, stressing that we have not found any  signature of a thermodynamic phase transition.


\section{Conclusions} 
\label{sec:discussion}

In this work, we numerically studied the glass-molten phase transition of RNA secondary structures using a Gaussian disorder model. With the guidance of previous studies of this system using renormalization-group theory~\cite{mlkw,fdkw_smallPaper,fdkw_bigPaper} we   determined the precise location of the transition by the introduction of two order parameters, one for the glass phase, and one for the molten phase. In addition to precisely locating the critical temperature of this system, we confirmed the numerical values of the critical exponents at the phase transition predicted by the field theory.

We  provide an explanation as to how this model transitions between its molten and glass states by studying the behavior of the base-pairing probability distributions. In particular, we show that the total base-pairing probability distribution can be broken into two sub-distributions composed of base pairs that are ``locked'' in the ground-state structures below the glass transition temperature, $Q_{\rm GS}(x;\ell)$, and non-ground-state base pairs, $Q_{\rm NGS}(x;\ell)$, that never develop a sizable pairing probability.  Interestingly, we also identify a potential {\em secondary freezing} event~\cite{mlkw} within the low temperature phase.  At temperatures below this secondary freezing event, the most prevalent base pairing probability of ground-state base pairs remains one even at finite temperatures indicating that they are truly locked.  Above this secondary freezing event but below the glass transition temperature, base pairs with significant pairing probabilities ($p\ge0.5$) remain prevalent but the density of base pairs with pairing probability one itself vanishes, indicating that locking is maintained only in the weaker sense that locked base pairs are common to a finite fraction of the thermodynamic ensemble rather than to ``all'' structures.  It would be   interesting to study the implications of this   secondary freezing event on the kinetics of RNA structures.

\section*{Acknowledgments}

This material is based upon work supported by the National Science Foundation under Grants No.\ DMR-1410172 and DMR-1719316.
K.W.\ thanks W.\ Janke for discussions and early simulations addressing the L\"assig-Wiese phase transition scenario, and for pointing out the importance of finite-size effects. We acknowledge useful discussions with F.\ David and M.\ L\"assig.

\appendix

\section{Construction of two order parameters in the two-dimensional Ising model}
\label{sec:ising}

\begin{figure*}\includegraphics[width=\columnwidth]{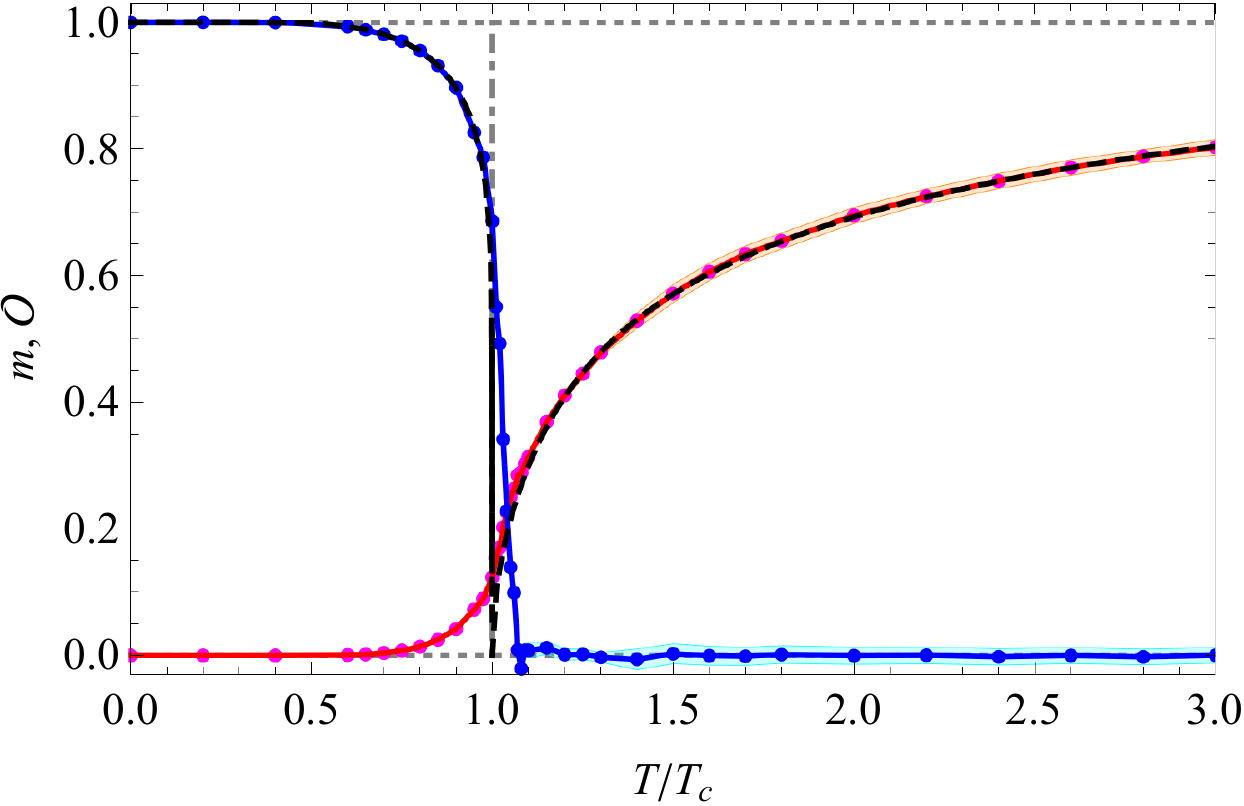}
\includegraphics[width=\columnwidth]{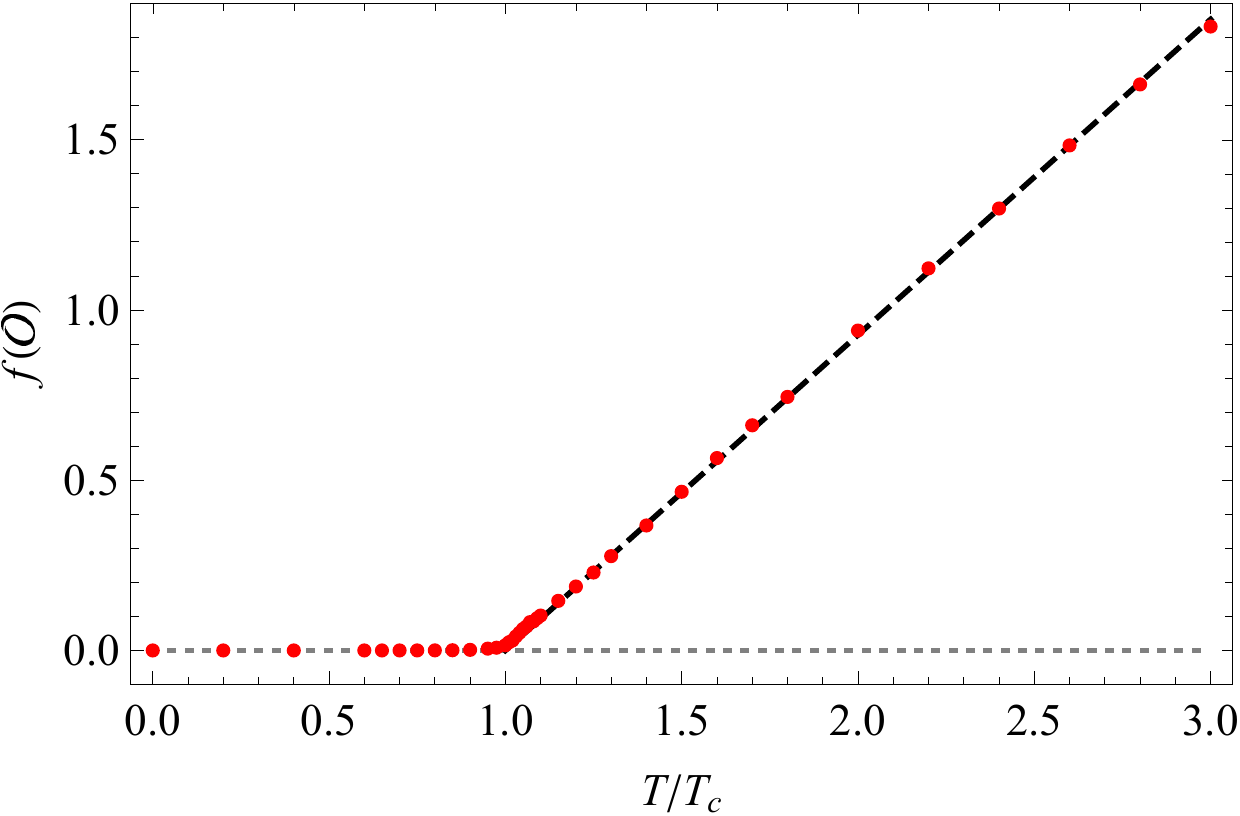}
\caption{Left: Simulation results for the two order parameters of the 2D Ising model at linear size $L=150$: Non-vanishing for $T<T_{\rm c}$ (blue in online version) the magnetization $m$; superimposed the analytical result of Yang \cite{Yang1952} (black dashed). Non-vanishing for $T>T_{\rm c}$ (red in online version), the high-temperature order-parameter $\cal O$  defined in Eq.~(\ref{Odef}). The black dashed line is a fit of the data,   starting  with a square-root singularity in $T-T_{\rm c}$ (exact form to be confirmed analytically). This can be verified on the right plot, where $f(x)=x^2+4 x^6 + 10 x^{20} $; the higher-order terms where chosen s.t.\ even for large $T/T_{\rm c}$ the data  roughly lie on a straight line, namely $0.927(T/T_{\rm c}-1)$. This shows that $\cal O$ is an order parameter for the high-temperature phase.
Simulations where performed for system size $L=150$, using a total of $1000 L^2$  updates per temperature in the region $1\le T\le 1.3$, and   $200 L^2$ updates elsewhere, starting from a configuration annealed at the next smaller temperature. Statistical fluctuations are given by the shaded regions. The dominant error is systematic.}
\label{f:Ising}
\end{figure*}

In this appendix we construct a pair of order parameters vanishing on either side of a phase transition in a system known to have only a single phase transition. To this end, consider the Curie point in magnets, separating a ferromagnetic from a paramagnetic phase. In the ferromagnetic phase, the magnetisation is an order parameter, vanishing when approaching  the transition from below. This order parameter has a $\mathbb Z_2$ symmetry, which is spontaneously broken in the ordered phase. The connected correlation function in the ordered phase decays exponentially, defining a correlation length $\xi_{\rm ferro}$, which diverges at the transition, $\xi _{\rm ferro}\sim |T-T_{\rm c}|^{-\nu^*}$.
While this correlation length does not show the $\mathbb Z_2$ symmetry breaking, its inverse could still  be used as an order parameter.

In the high-temperature phase
it is more difficult to find an order parameter. One idea is   to use the (inverse) correlation length, which vanishes as $1/\xi _{\rm para}\sim |T-T_{\rm c}|^{\nu^*}$. A better observable can be constructed as follows: Draw a line along one of the lattice directions, and note the numbers $S$ of consecutive spins with the same sign.
As an example, in a system of linear size $L=10$ with periodic boundary conditions, the configuration ``0111100010'' on the chosen line translates into the 4 random variables $\{ S_1,...,S_{4}\}=\{4,3,1,2\}$. 
  Define the observable 
\begin{equation}\label{Odef}
{\cal O} := \left( \frac {\left<S \right>}{\left< S^2 \right>}-\frac1{L}\right)\times \left( \frac13-\frac1L \right)^{\!-1}\ .
\end{equation}
At $T=0$, all spins are aligned and $S$ takes only one value, $S=L$. By construction $\cal O$ vanishes. For $T\to \infty$, the spins are independent, and the ratio $ {\left<S \right>}/{\left< S^2 \right>}\to 1/3$. Thus $\cal O$ approaches 1 for $T\to \infty$. The behavior for intermediate temperatures is shown in Fig.~\ref{f:Ising} (left):
The observable $\cal O$ vanishes (approximately) for  $T\le T_{\rm c}$ and roughly grows as ${\cal O} \simeq 0.927 \sqrt{T-T_{\rm c}}$ for $T>T_{\rm c}$. This is more clearly seen by   plotting  ${\cal O}^2$ (plus higher-order terms to bring the result closer to a straight line for larger $T$), see the right of Fig.\ \ref{f:Ising}. A small rounding around $T=T_{\rm c}$ persists due to finite-size effects. 
Since there is only one phase transition in the 2D Ising model, this shows that one can construct two order parameters for this single transition.

\end{document}